\documentclass{article}

\PassOptionsToPackage{numbers, compress}{natbib}
\usepackage[preprint]{neurips_2026}

\usepackage[utf8]{inputenc} 
\usepackage[T1]{fontenc}    
\usepackage{hyperref}       
\usepackage{url}            
\usepackage{booktabs}       
\usepackage{amsfonts}       
\usepackage{nicefrac}       
\usepackage{microtype}      
\usepackage{xcolor}         

\usepackage{booktabs} 
\usepackage{booktabs}  
\usepackage{pifont}    
\usepackage{graphicx} 
\usepackage{multirow} 
\usepackage{tabularx} 
\usepackage[most]{tcolorbox}
\tcbuselibrary{skins} 
\tcbuselibrary{breakable}
\usepackage{enumitem}
\usepackage{graphicx}
\usepackage{subcaption}
\usepackage{caption} 
\PassOptionsToPackage{table,xcdraw}{xcolor}
\newcommand{\cmark}{\textcolor{teal}{\ding{51}}} 
\newcommand{\xmark}{\textcolor{lightgray}{\ding{55}}} 

\usepackage{listings}

\colorlet{punct}{red!60!black}
\definecolor{background}{HTML}{EEEEEE}
\definecolor{delim}{RGB}{20,105,176}
\colorlet{numb}{magenta!60!black}

\lstdefinelanguage{json}{
    basicstyle=\normalfont\ttfamily,
    numbers=left,
    numberstyle=\scriptsize,
    stepnumber=1,
    numbersep=8pt,
    showstringspaces=false,
    breaklines=true,
    frame=lines,
    backgroundcolor=\color{background},
    literate=
     *{0}{{{\color{numb}0}}}{1}
      {1}{{{\color{numb}1}}}{1}
      {2}{{{\color{numb}2}}}{1}
      {3}{{{\color{numb}3}}}{1}
      {4}{{{\color{numb}4}}}{1}
      {5}{{{\color{numb}5}}}{1}
      {6}{{{\color{numb}6}}}{1}
      {7}{{{\color{numb}7}}}{1}
      {8}{{{\color{numb}8}}}{1}
      {9}{{{\color{numb}9}}}{1}
      {:}{{{\color{punct}{:}}}}{1}
      {,}{{{\color{punct}{,}}}}{1}
      {\{}{{{\color{delim}{\{}}}}{1}
      {\}}{{{\color{delim}{\}}}}}{1}
      {[}{{{\color{delim}{[}}}}{1}
      {]}{{{\color{delim}{]}}}}{1},
}

\title{FinDocMRE: A Benchmark for Document-Level Financial Multimodal Reasoning Evaluation}

\author{
  Jiayong Zhu$^{1}$, \ \  
  Jiangtong Li$^{1}$, \ \
  Jinru Ding$^{2}$, \ \
  Dawei Cheng$^{1}$, \ \ 
  Jie Xu$^{2}$, \ \ 
  Feng Yu$^{3}$ \\ \\
  \small{$1.$ School of Computer Science and Technology, Tongji University, Shanghai, China} \\
  \small{$2.$ Shanghai Artificial Intelligence Laboratory, Shanghai, China }\\
  \small{$3.$ Guotai Haitong Securities Co.,Ltd. }\\
  \texttt{\{jiayongz, jiangtongli, dcheng\}@tongji.edu.cn} \\
}

\begin{document}

\maketitle

\begin{abstract}
While Large Multimodal Models~(LMMs) excel in general visual tasks, their deployment in specialized financial contexts remains insufficient.
Existing benchmarks prioritize isolated charts, often overlooking the need to integrate data from text, tables, and images within comprehensive financial documents.
To address this limitation, we introduce \textsc{\textbf{FinDocMRE}}, a multi-image document-level benchmark designed for financial multimodal reasoning.
We construct the dataset via a semi-automated pipeline that combines Visual-Centric Generation with Expert Verification, thereby minimizing text bias and ensuring high annotation quality.
Spanning twelve domains, the benchmark comprises 12,207 samples derived from 2,878 financial reports, designed to evaluate multi-image processing and document-level understanding across five distinct task types.
Extensive experiments with eleven representative LMMs reveal that no model surpasses an overall score of 65, highlighting challenges in integrating visual grounding with logical reasoning within complex document environments.
Specifically, we observe a significant performance divergence across tasks, where models exhibit proficiency in semantic narrative construction but struggle with numerical estimation and cross-page visual grounding.
\textsc{FinDocMRE} serves as a rigorous benchmark to guide the evolution of financial LMMs towards expert-level document analysis and reasoning.
\end{abstract}

\section{Introduction}


Recent progress in Large Multimodal Models~(LMMs) has significantly improved multimodal understanding and reasoning capabilities~\cite{comanici2025gemini,bai2025qwen2}.
Therefore, applying these models within the financial domain has attracted growing interest.
As LMMs evolve from general assistants into specialized financial agents, rigorous evaluation benchmarks become essential.
These benchmarks serve as metrics for progress while guiding model development to meet the demands of professional analysis.


\begin{figure*}[t] 
    \centering
    \includegraphics[width=\textwidth]{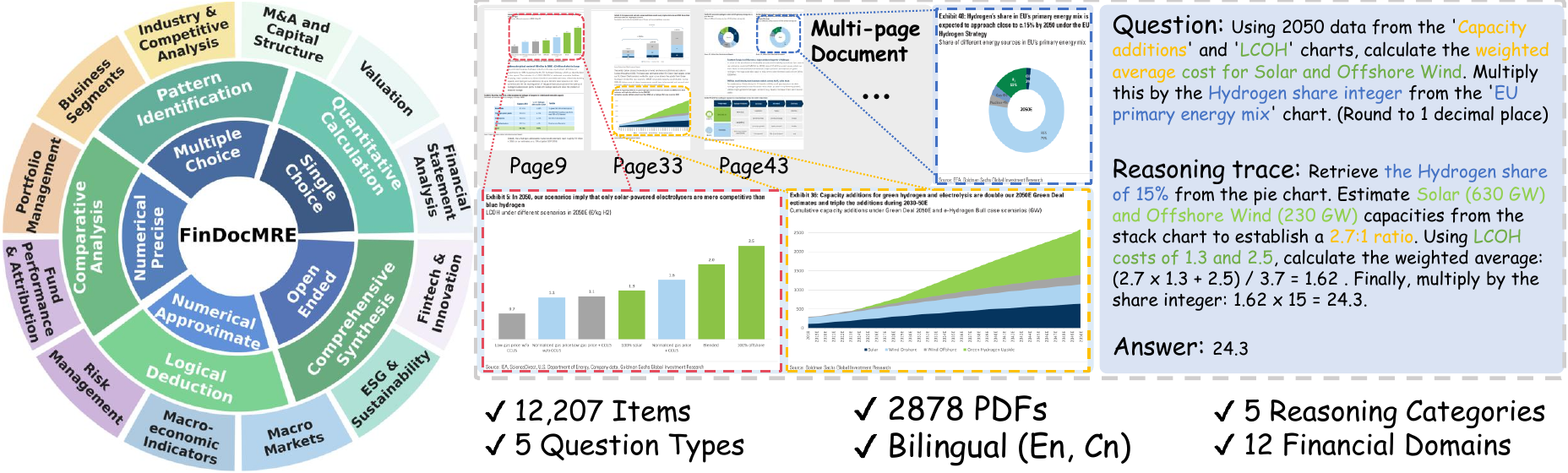} 
    \caption{The overall taxonomy and statistics of the \textsc{FinDocMRE} benchmark. An example illustrates the organization of a multi-image document-level sample.}
    \label{fig:overview}
\end{figure*}

Previous research has established various methods for assessing AI models' financial capabilities.
However, existing benchmarks primarily target natural language processing~(NLP) tasks~\cite{chen2021finqa, ding2025cnfinbench} or isolated chart comprehension~\cite{xue2024famma}.
Consequently, these datasets fail to adequately represent the complexity of real-world \textbf{document-level} financial analysis.
In contrast to isolated tasks, document-level analysis involves integrating interwoven text, dense data tables, and visual charts, often requiring cross-page reasoning.
Despite such evaluation being necessary, constructing a document-level benchmark presents three primary challenges:
1) \textbf{Data Availability:} High-quality, open-source financial PDFs providing structural alignments are rare compared to plain text corpora, limiting multimodal evaluation.
2) \textbf{Annotation Costs:} Manually designing complex, multi-hop reasoning questions from full documents is highly labor-intensive and demands substantial domain expertise, limiting both dataset scale and diversity.
3) \textbf{Generation Quality:} While generating questions via LMMs is scalable, prior work indicates this method introduces \textbf{text bias}~\cite{karamcheti2024prismatic,huang2024visual}, yielding coarse-grained queries that lack the numerical precision required for testing expert-level visual reasoning.
 
To address these challenges, we introduce \textsc{\textbf{FinDocMRE}} (\textbf{Fin}ancial \textbf{Doc}ument-level \textbf{M}ultimodal \textbf{R}easoning \textbf{E}valuation), a benchmark designed to assess LMMs within document-level financial reasoning scenarios.
We establish a semi-automated construction pipeline to balance scalability and data quality:
1) \textbf{Data Preparation:} We collect bilingual financial PDF reports (\emph{e.g.}, annual and research reports) from our collaborators and apply a Joint-Rule Filtering Mechanism (geometric, semantic, and textual indexing) to extract high-value charts and remove noise;
2) \textbf{Visual-Centric QA Generation:} We input aggregated chart contexts (excluding surrounding text) into LMMs to ensure the generation of precise, vision-dependent QA pairs and reasoning traces;
3) \textbf{Expert Verification:} We employ a human-in-the-loop protocol where senior financial experts verify visual grounding~(chart IDs) and correct logical errors, retaining only valid samples for the final dataset.
As detailed in Tab.~\ref{tab:comparison}, \textsc{FinDocMRE} stands out as a large-scale financial benchmark supporting \textbf{multi-image document-level} inputs, featuring the largest data scale and diverse answer formats.
As shown in Fig.~\ref{fig:overview}, the benchmark comprises 12,207 samples derived from 2,878 financial reports spanning twelve industries, covering five question formats and five reasoning categories.
We conduct comprehensive evaluations of eleven representative LMMs alongside human expert baselines on \textsc{FinDocMRE}, employing full PDF formats to simulate realistic financial workflows.
Our multidimensional analysis reveals a significant divergence between quantitative calculation and qualitative synthesis: \textbf{while models excel at semantic narrative construction, they struggle with precise trend estimation}.
Notably, performance deteriorates as visual dependency increases, underscoring that multi-image reasoning remains a bottleneck for current leading multimodal architectures.

To investigate these limitations, we perform multidimensional analyses and ablation studies exploring factors like visual context, model scaling, visual complexity, and prompt robustness, among others.
Reducing the visual search space leads to substantial performance gains, indicating a deficiency in aggregating fragmented information from extensive document contexts.
With no model surpassing an overall score of 65, our results highlight the formidable challenge of integrating visual grounding with rigorous numerical reasoning.
In summary, our contributions are as follows:
\begin{itemize}
    \item We present \textsc{FinDocMRE}, the first large-scale benchmark addressing multi-image document-level financial multimodal reasoning, filling the gap in current evaluation standards.
    \item We develop a semi-automated pipeline integrating Visual-Centric Generation with Expert Verification, reducing textual noise to yield 12,207 high-quality QA samples from 2,878 reports across twelve financial domains.
    \item We conduct a multidimensional evaluation of eleven representative LMMs against real-world financial expert baselines, employing comparative analysis (Full PDF vs. Cropped) to isolate specific deficiencies in visual grounding and quantitative reasoning.
\end{itemize}

\begin{table}[t!]
    \centering
    \caption{Comparison with representative benchmarks. \textbf{Vol.}: Volume; \textbf{MM}: Multimodal; \textbf{Fin.}: Financial; \textbf{M-Img}: Multi-image; \textbf{Doc.}: Document-level; \textbf{MC}: Multiple-choice; \textbf{N$_{pre}$}: Numerical precise; \textbf{N$_{app}$}: Numerical approximate; \textbf{Open}: Open-ended QA.} 
    \label{tab:comparison}
    \small 
    
    \setlength{\tabcolsep}{8pt} 
    \begin{tabular}{l ccccc cccc}
        \toprule
        & \multicolumn{5}{c}{\textbf{Dataset Features}} & \multicolumn{4}{c}{\textbf{Answer Format}} \\
        \cmidrule(lr){2-6} \cmidrule(lr){7-10}
        \textbf{Dataset} & \textbf{Vol.} & \textbf{MM} & \textbf{Fin.} & \textbf{M-Img} & \textbf{Doc.} & \textbf{MC} & \textbf{N$_{pre}$} & \textbf{N$_{app}$} & \textbf{Open} \\
        \midrule
        MMStar       & 1,500  & \cmark & \xmark & \xmark & \xmark & \xmark & \xmark & \xmark & \xmark \\
        MathVision   & 3,040  & \cmark & \xmark & \xmark & \xmark & \xmark & \cmark & \xmark & \xmark \\
        \midrule
        FinEval      & 8,351  & \xmark & \cmark & \xmark & \xmark & \xmark & \cmark & \xmark & \cmark \\
        FinQA        & 8,281  & \xmark & \cmark & \xmark & \xmark & \xmark & \cmark & \xmark & \xmark \\
        FinanceBench & 10,231 & \xmark & \cmark & \xmark & \xmark & \xmark & \cmark & \xmark & \cmark \\
        \midrule
        FAMMA        & 1,935  & \cmark & \cmark & \cmark & \xmark & \xmark & \cmark & \xmark & \cmark \\
        MME-Finance  & 2,274  & \cmark & \cmark & \xmark & \xmark & \xmark & \cmark & \cmark & \cmark \\
        FinMME       & 11,099 & \cmark & \cmark & \xmark & \xmark & \cmark & \cmark & \xmark & \xmark \\
        FinMMDocR    & 1,200  & \cmark & \cmark & \cmark & \cmark & \xmark & \cmark & \xmark & \xmark \\
        \midrule
        \textbf{FinDocMRE} & \textbf{12,207} & \textbf{\cmark} & \textbf{\cmark} & \textbf{\cmark} & \textbf{\cmark} & \textbf{\cmark} & \textbf{\cmark} & \textbf{\cmark} & \textbf{\cmark} \\
        \bottomrule
    \end{tabular}
\end{table}

\section{Related Work}
\subsection{Large Multi-modal Models}
Significant advancements in LMMs, ranging from proprietary models like GPT~\cite{achiam2023gpt}, Gemini~\cite{comanici2025gemini}, Claude~\cite{anthropic2024claude3}, GLM~\cite{zeng2025glm}, and SEED~\cite{guo2025seed1}, to open-source counterparts like LLaVA~\cite{liu2023visual}, Qwen-VL~\cite{bai2025qwen2}, and InternVL~\cite{zhu2025internvl3}, have greatly enhanced general visual understanding and multimodal reasoning.
However, deploying these general-purpose models in the financial domain faces distinct challenges, as professional analysis demands not only visual recognition but also precise numerical reasoning and trend interpretation within noise-sensitive contexts.
Therefore, while these foundation models demonstrate broad capabilities, their reliability as financial agents in realistic, document-intensive scenarios remains under-explored, necessitating evaluation paradigms that go beyond isolated tasks.

\subsection{Multimodal Benchmarks}
To assess the evolving capabilities of LMMs, various general-purpose multimodal benchmarks have been introduced, including MME~\cite{fu2023mme}, MathVision~\cite{wang2024measuring}, MathVerse~\cite{zhang2024mathverse}, MMMU~\cite{yue2024mmmu}, MathVista~\cite{lu2023mathvista}, ChartQA~\cite{masry2022chartqa}, MM-Star~\cite{chen2024we}, and others~\cite{liu2024mmbench,xu2023chartbench,liu2024mmc}.
However, these frameworks often fall short in specialized financial contexts, typically confining tasks to isolated visual question answering and omitting the complex grounding demands of the domain. 
In parallel, document-centric datasets like DocVQA~\cite{mathew2021docvqa} heavily emphasize Optical Character Recognition~(OCR) and entity extraction, often bypassing the high-level reasoning required to synthesize cross-modal data~(text, tables, and charts). 
These limitations motivate the development of \textsc{FinDocMRE}, which aims to bridge basic visual perception with expert-level document analysis and reasoning.

\subsection{Benchmarks in Finance}
In the text domain, benchmarks have evolved from focusing on specific NLP and reasoning tasks, exemplified by FLUE~\cite{shah2022flue}, CFLEB~\cite{lu2023bbt}, InvestorBench~\cite{li2024investorbench}, and FinanceBench~\cite{islam2023financebench}, to evaluating holistic financial literacy, as seen in suites like FinEval~\cite{zhang2023fineval}, FinBen~\cite{xie2024finben}, CFBenchmark~\cite{lei2023cfbenchmark}, SuperCLUE-Fin~\cite{xu2024superclue}, OpenFinData~\cite{openfindata}, and CNFinBench~\cite{ding2025cnfinbench}.
More recently, initiatives like CFBenchmark-MM~\cite{li2025cfbenchmark}, MME-Finance~\cite{gan2025mme}, FinMME~\cite{luo2025finmme}, FinMR~\cite{deng2025finmr}, FinReasoning~\cite{zhu2026comprehension}, and FinTBS~\cite{hu2025fintsb} have extended evaluation to the visual modality, targeting chart and table interpretation.
Nevertheless, these datasets typically simplify the task by relying on isolated, pre-localized images.
The concurrent work FinMMDocR~\cite{tang2026finmmdocr} investigates document-level financial reasoning, focusing on numerical computation. While sharing this setting, \textsc{FinDocMRE} provides a larger-scale evaluation and introduces more diverse task formats, including multiple-choice and open-ended QA, demanding a more comprehensive understanding of complex financial logic and holistic document intelligence.


\begin{figure*}[t] 
    \centering
    \includegraphics[width=\textwidth]{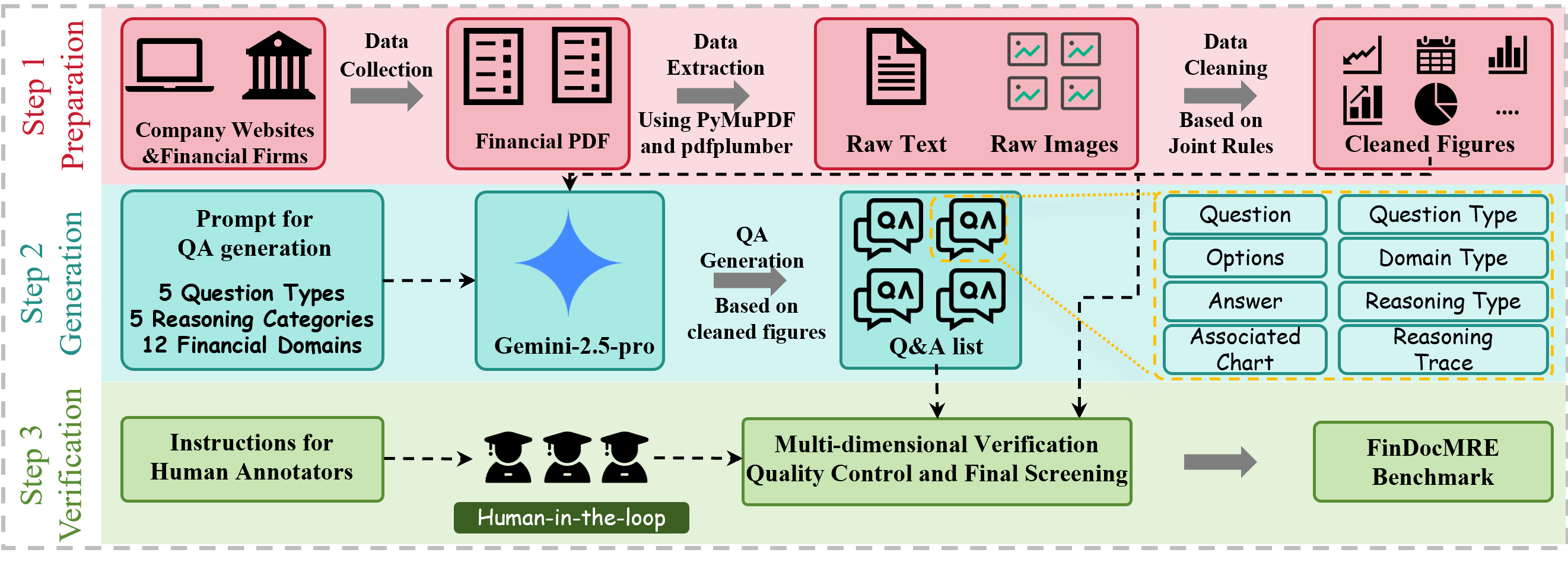} 
    \caption{The annotation pipeline of \textsc{FinDocMRE} benchmark .}
    \label{fig:pipeline}
\end{figure*}

\section{\textsc{FinDocMRE} Benchmark}
As illustrated in Fig.~\ref{fig:pipeline}, \textsc{FinDocMRE} is constructed through a semi-automated pipeline: Data Preparation~(Sec.~\ref{sec:bench:pre}), Visual-Centric Generation~(Sec.~\ref{sec:bench:gen}), and Expert Verification~(Sec.~\ref{sec:bench:ver}).
This framework combines the scalability of automated generation with strict human-in-the-loop protocols to ensure the accuracy of financial reasoning.
Comprehensive statistics are in Sec.~\ref{sec:bench:stat}.

\subsection{Data Preparation}\label{sec:bench:pre}
\noindent\textbf{Data Acquisition.}
To construct a benchmark reflecting real-world scenarios, we collect a corpus of 2,878 financial PDFs~(2,469 Chinese, 409 English) from official websites and third-party aggregators.
These sources encompass authoritative domestic and international institutions, spanning leading securities firms and investment banks to global economic organizations.
We prioritize document types essential for professional analysis, including annual reports, weekly newsletters, and industry research reports, while excluding plain-text files to explicitly target multimodal reasoning.

\noindent\textbf{Chart-Centric Extraction and Cleaning.}
To process these unstructured documents, we implement a Chart-Centric extraction pipeline, using PyMuPDF and pdfplumber to parse embedded images and textual content.
To maintain data quality, we apply a Joint-Rule Filtering Mechanism: (1) Geometric Filtering discards layout artifacts based on extreme aspect ratios or low resolutions; (2) Semantic Filtering utilizes image similarity computation to deduplicate visual content, while employing OCR to filter out non-informative elements like corporate logos; and (3) Textual Indexing Verification retains only charts explicitly referenced in the narrative (\emph{e.g.}, ``Figure 1'') to ensure analytical relevance.
\textbf{Further details on data extraction, cleaning and annotation are provided in the Appendix~\ref{app:benchmark_construction}.}

\subsection{Visual-Centric Generation}\label{sec:bench:gen}

Preliminary experiments reveal that feeding full PDF pages or images with surrounding text introduces textual bias, compromises accurate chart interpretation.
To mitigate this, we employ a visual-centric generation strategy using \texttt{Gemini-2.5-Pro}.
As shown in Fig.~\ref{fig:pipeline}-Step 2, for each document, we combine all extracted charts into an image sequence and prompt~(Tab.~\ref{tab:prompt_full}) the model to generate diverse question-answer pairs and reasoning traces, prioritizing cross-chart synthesis where applicable.
Excluding surrounding text, we compel the model to derive insights solely from visual data, simulating the workflow of an analyst where visual evidence is interpreted independently of textual summaries.

To ensure a gradient of difficulty levels, we incorporate a multi-dimensional taxonomy directly into the generation prompts.
We apply strict JSON schema constraints across five question formats: \textit{\textbf{single\_choice}} and \textit{\textbf{multiple\_choice}} tasks require plausible distractors; \textit{\textbf{numerical\_precise}} and \textit{\textbf{numerical\_approximate}} distinguish exact results from trend estimation; and \textit{\textbf{open\_ended}} queries require synthesizing financial concepts.
As illustrated in Fig.~\ref{fig:overview}, this taxonomy spans 12 financial domains (\emph{e.g.}, \textit{Financial Statement Analysis}, \textit{Risk Management}) and five cognitive reasoning categories, progressing from basic \textit{Quantitative Calculation} to advanced \textit{Comprehensive Synthesis}.

Each generated sample is structured as a metadata object containing the \textit{Question}, \textit{Answer}, \textit{Domain Type}, and \textit{Reasoning Type}.
We also require the generation of a detailed \textit{Reasoning Trace}, outlining the step-by-step logic and calculation formulas used to derive the answer.
This trace grounds the model inference and assists the subsequent Expert Verification phase, allowing annotators to validate the correctness of complex derivations.
Finally, the model records specific \texttt{chart\_ids} in the \textit{Associated Chart} field, linking each query to its source evidence within the multi-image input.

\subsection{Expert Verification}\label{sec:bench:ver}
To counteract the hallucination risks inherent to visual-centric generation and guarantee expert-level fidelity, we deploy a Human-In-The-Loop (HITL) verification workflow (Fig.~\ref{fig:pipeline}-Step~3).
Each document and its corresponding generated sample are verified by three senior financial researchers, each possessing over three years of professional experience in securities firms.
Guided by rigorous annotation protocols in Tab.~\ref{tab:human_guidelines}, experts audit each sample by cross-referencing the generated reasoning traces against the source charts.
This examination validates the logic of \textbf{Question}, the plausibility of \textbf{Options}, the accuracy of \textbf{Answer}, and the precision of \textbf{Associated Chart} IDs.

To ensure the highest data quality, the verification team performs strict filtering rather than manual correction.
During the audit, researchers explicitly reject any sample that displays ambiguous phrasing, factual errors, weak option quality, or mismatched chart references.
We adopt a unanimous consensus mechanism: if any of the three experts identifies a flaw, the sample is immediately discarded.
Therefore, only samples validated by all three researchers without reservation are retained in the final \textsc{FinDocMRE} benchmark.
This rigorous quality control ensures adherence to professional analysis standards, resulting in a final retention rate of approximately 55\% for \textsc{FinDocMRE}.

\begin{figure}[t!]
    \centering
    \begin{minipage}[b]{0.38\linewidth}
        \centering
        \captionof{table}{Statistics of the FinDocMRE dataset.}
        \resizebox{\linewidth}{!}{
            \begin{tabular}{lc}
                \toprule
                \textbf{Category} & \textbf{Count} \\
                \midrule
                \textit{Questions (QA Pairs)} & \\ 
                \cmidrule(r){1-2}                
                Total Questions & 12,207 \\
                \quad -- Chinese & 7,105 \\
                \quad -- English & 5,102 \\
                \midrule 
                \textit{Source Documents (PDFs)} & \\
                \cmidrule(r){1-2}
                Total PDFs & 2,878 \\
                \quad -- Chinese & 2,469 \\
                \quad -- English & 409 \\
                \midrule
                \textit{Question Types Distribution} & \\
                \cmidrule(r){1-2}
                Single Choice & 3,439 \\
                \cmidrule(r){1-2}
                Multiple Choice & 1,423 \\
                \cmidrule(r){1-2}
                Numerical Precise & 2,405 \\
                \cmidrule(r){1-2}
                Numerical Approximate & 2,454 \\
                \cmidrule(r){1-2}
                Open Ended & 2,486 \\
                \bottomrule
            \end{tabular}
        }
        \label{tab:statistics}
    \end{minipage}
    \hfill
    \begin{minipage}[b]{0.58\linewidth}
        \centering
        \begin{subfigure}[b]{0.35\linewidth}
            \centering
            \includegraphics[width=\linewidth]{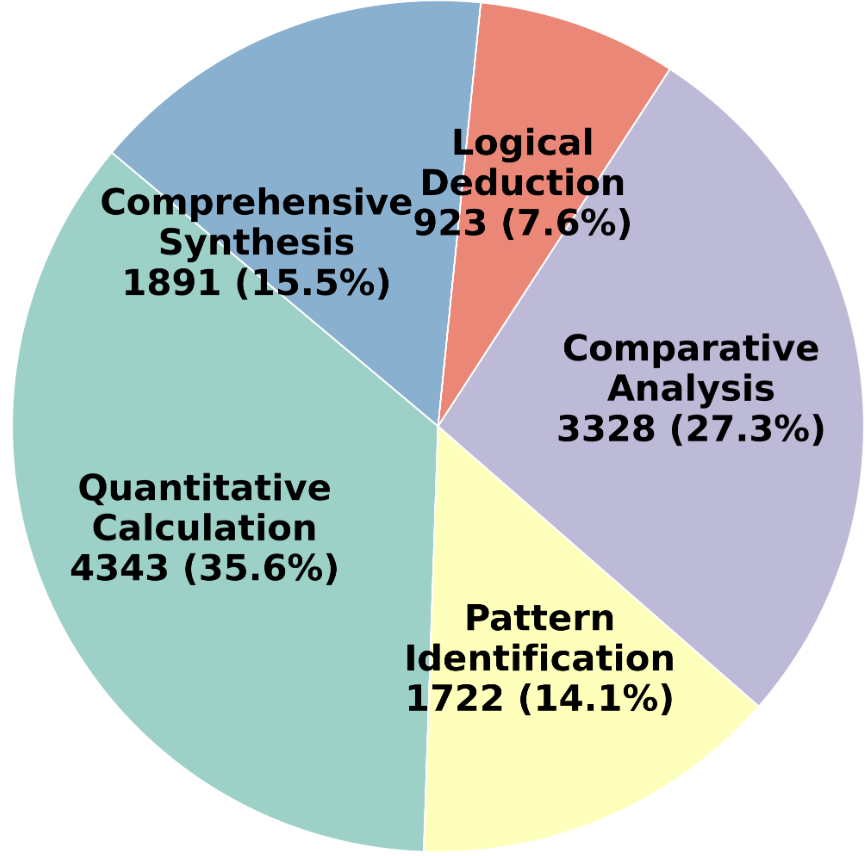}
            \caption{}
            \label{fig:reasoning}
        \end{subfigure}
        \hfill
        \begin{subfigure}[b]{0.62\linewidth}
            \centering
            \includegraphics[width=\linewidth]{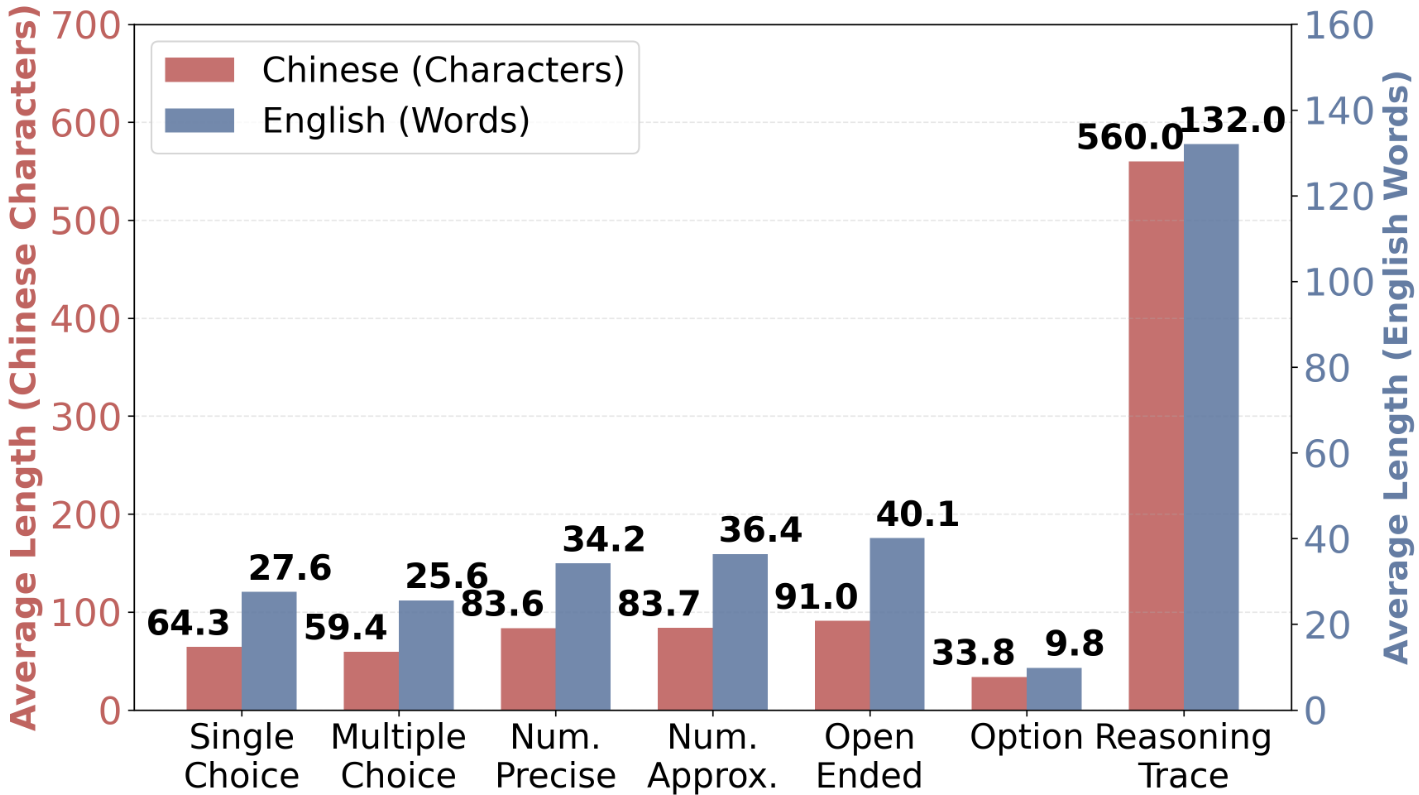}
            \caption{}
            \label{fig:length}
        \end{subfigure}
        
        
        \begin{subfigure}[b]{0.52\linewidth}
            \centering
            \includegraphics[width=\linewidth]{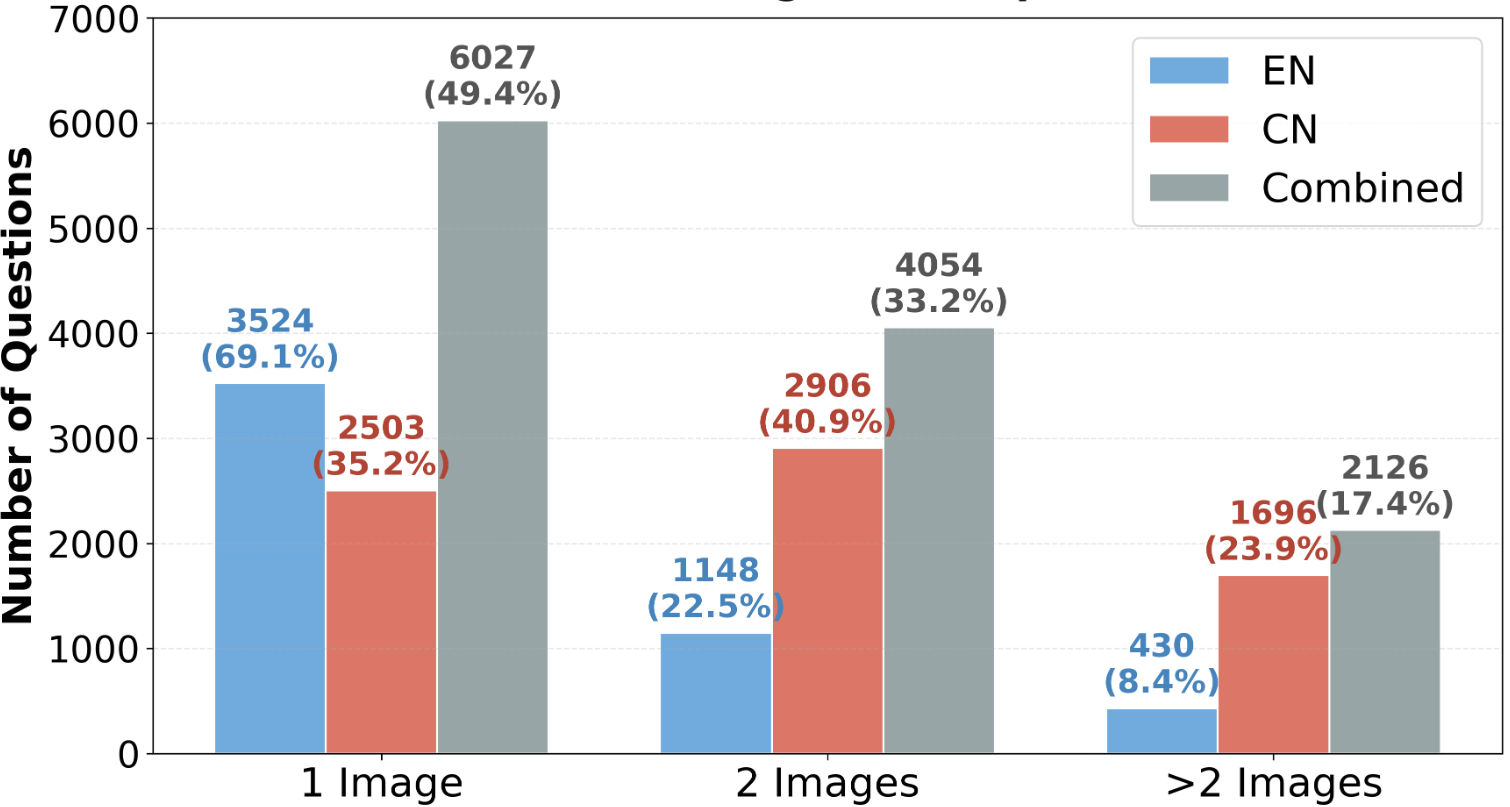}
            \caption{}
            \label{fig:size}
        \end{subfigure}
        \hfill
        \begin{subfigure}[b]{0.44\linewidth}
            \centering
            \includegraphics[width=\linewidth]{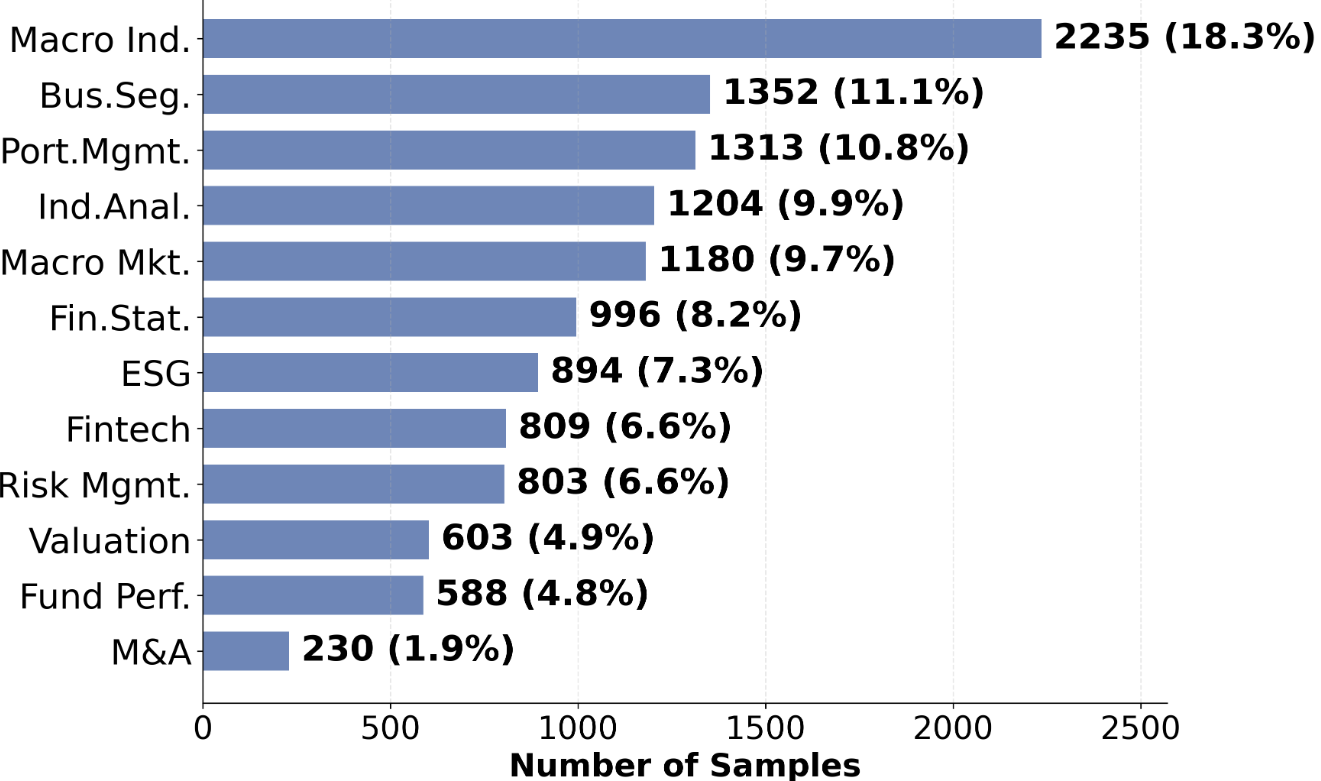}
            \caption{}
            \label{fig:domain}
        \end{subfigure}
        \captionof{figure}{Dataset statistics overview: (a) Reasoning types, (b) Lengths, (c) Image quantity, (d) Domains.}
        \label{fig:dataset_stats_custom_ratio}
    \end{minipage}
\end{figure}

\subsection{Dataset Statistics}\label{sec:bench:stat}

\noindent\textbf{Scale and Linguistic Diversity.}
\textsc{FinDocMRE} contains 12,207 annotated samples derived from 2,878 unique financial documents, comprising 7,105 Chinese and 5,102 English QA pairs, carefully balanced to evaluate cross-lingual reasoning capabilities.
Rooted in native reports from major global markets, this bilingual composition ensures robust applicability across diverse financial contexts.

\noindent\textbf{Distribution of Question Types.}
In contrast to general multimodal datasets focused on basic visual recognition tasks, \textsc{FinDocMRE} prioritizes quantitative analysis.
As shown in Tab.~\ref{tab:statistics}, \textit{Numerical Reasoning} represents a core component: \textit{numerical\_precise} (2,405) and \textit{numerical\_approximate} (2,454) together account for 39.8\% of the total samples.
Furthermore, \textit{open\_ended} queries (2,486) represent 20.4\% of the total, demanding coherent textual synthesis beyond mere option selection.

\noindent\textbf{Reasoning Depth and Visual Dependency.}
To assess reasoning difficulty, we examine the length of the expert-verified Chain-of-Thought~(CoT) sequences.
The average CoT length is 560 characters for Chinese and 132 words for English samples, implying the necessity of multi-step deduction to resolve complex financial queries.
Regarding visual grounding, approximately 50.6\% of the questions require synthesizing information across multiple charts, necessitating cross-page reasoning within extensive document contexts.
Fig.~\ref{fig:dataset_stats_custom_ratio} details the distribution across 12 financial domains and 5 reasoning types, illustrating the coverage of \textsc{FinDocMRE} across different analysis scenarios.

    
    
    
    
    
    
    

\begin{table*}[t]
\centering
\caption{Main evaluation results on \textsc{FinDocMRE}.
We report Accuracy (\%) for objective tasks and raw 0-5 scores for open-ended tasks (normalized percentages in parentheses).
The Overall Score is the macro-average of the five metrics.
The best model results in each column are marked in \textbf{bold}.}
\resizebox{\textwidth}{!}{
\setlength{\tabcolsep}{10pt}
\begin{tabular}{l|ccccc|c}
\toprule
\textbf{Model} & \textbf{Single-Ch.} & \textbf{Multi-Ch.} & \textbf{Num. (Pre.)} & \textbf{Num. (App.)} & \textbf{Open-Ended} & \textbf{Overall Score} \\
\midrule
 & \multicolumn{6}{c}{\textit{Proprietary Models}} \\ \midrule
Gemini-3.1 Pro & \textbf{79.12} & \textbf{58.05} & \textbf{63.20} & \textbf{37.82} & 4.12 (82.4) & \textbf{64.12} \\
GPT-5.4 & 76.04 & 56.71 & 57.17 & 31.21 & \textbf{4.41} (88.2) & 61.87 \\
Doubao-1.6 Vision & 73.77 & 57.62 & 50.31 & 26.65 & 4.19 (83.8) & 58.43 \\
GPT-5 & 73.02 & 51.16 & 48.07 & 26.73 & 4.33 (86.6) & 57.12 \\
Gemini-2.5 Pro & 72.70 & 45.33 & 36.22 & 20.70 & 3.90 (78.0) & 50.59 \\
Gemini-2.5 Flash & 64.55 & 41.67 & 25.16 & 20.95 & 3.94 (78.8) & 46.23 \\
Qwen3-Max & 49.96 & 27.69 & 3.45 & 11.86 & 3.51 (70.2) & 32.63 \\
Grok-4.1 Fast & 53.97 & 23.19 & 3.74 & 9.90 & 3.26 (65.2) & 31.20 \\
\midrule
 & \multicolumn{6}{c}{\textit{Open-Source Models}} \\ \midrule
Qwen3-VL-235B & 75.25 & 52.64 & 52.89 & 26.28 & 4.08 (81.6) & 57.73 \\
Qwen3-VL-30B & 68.94 & 43.43 & 41.16 & 20.70 & 3.74 (74.8) & 49.81 \\
Qwen2.5-VL-72B & 59.44 & 28.95 & 14.10 & 14.43 & 3.19 (63.8) & 36.14 \\
\midrule
 & \multicolumn{6}{c}{\textit{Human Performance}} \\ \midrule
 Human Expert & 88.37 & 76.04 & 85.03 & 62.59 & 4.25 (85.0) & 79.41 \\
\bottomrule
\end{tabular}
}

\label{tab:main_results}
\end{table*}

\section{Experiment}

\subsection{Evaluated LMMs}
To assess the performance of state-of-the-art LMMs on \textsc{FinDocMRE}, we select 9 models ranging from commercial APIs to open-weights architectures.
Our evaluation includes nine \textbf{proprietary models}: Gemini-3.1-Pro, GPT-5.4, GPT-5~\cite{GPT-5}, Doubao1.6-Vision~\cite{guo2025seed1}, Qwen3-Max~\cite{yang2025qwen3}, Grok4.1-Fast~\cite{grok-4-1}, Gemini-2.5-Pro, and Gemini-2.5-Flash~\cite{comanici2025gemini}.
We also evaluate three \textbf{open-source models}: Qwen2.5-VL-72B~\cite{bai2025qwen2}, Qwen3-VL-30B-A3B, and Qwen3-VL-235B-A22B~\cite{yang2025qwen3}.
All models are tested in a zero-shot setting to assess their inherent financial document reasoning capabilities without external tools.
Our preliminary screening excludes smaller architectures (\emph{e.g.}, Qwen2.5-VL-3B/7B) due to limited capabilities, and models like GLM and DeepSeek-VL, where safety filters cause high refusal rates on financial documents.
Finally, to establish a human performance baseline, we engage two professional financial analysts from securities firms to complete the benchmark under identical input conditions, utilizing their averaged scores to represent the \textbf{Human Expert} results.

\subsection{Evaluation Metrics}
We employ a dual-evaluation strategy aligned with specific task requirements.
For deterministic tasks (single-choice, multiple-choice, and numerical reasoning), we report standard \textbf{Accuracy}.
We apply strict validation protocols: multiple-choice questions require an \textbf{Exact Match} to the ground truth option set; numerical estimation allows a \textbf{5\% relative error tolerance} for visual ambiguity, while precise calculations require exact equality.
For open-ended queries, where n-gram metrics (\emph{e.g.}, BLEU~\cite{papineni2002bleu}) fail to capture semantic validity, we adopt an LLM-as-a-Judge paradigm~\cite{li2025generation}.
To mitigate single-evaluator bias, we employ a panel of three distinct models (\texttt{Gemini-2.5-Flash}, \texttt{GPT-4o}, and \texttt{Qwen3-Max}) to score responses against reference answers from [0,5], assessing correctness, completeness, and logical coherence, taking their average as the final score.
\textbf{See Appendix~\ref{app:bias} for judge  bias and alignment analysis between LMM evaluators and human experts.}

\noindent\textbf{Overall Performance Metric.}
To mitigate sample size imbalances across task categories, we calculate the \textbf{Overall Score} using a macro-average strategy.
We scale \textit{open\_ended} scores to a 0-100 range ($S_{norm} = S_{raw} \times 20$) and compute the unweighted mean across all five question types.
This method reduces bias from majority classes, providing a balanced assessment of reasoning abilities across dimensions.
\textbf{Refer to Appendix~\ref{app:evaluation_protocol} for full evaluation setting and prompts.}

\begin{table*}[t!]
\centering
\setlength{\tabcolsep}{14pt}
\caption{Performance breakdown across five reasoning categories. Model performance is formatted as Accuracy (\%) / Score (0-5). Note that the \textbf{Quant. Calc.} category consists exclusively of objective questions, so only Accuracy is reported.  Best results are marked in \textbf{bold}.}
\label{tab:reasoning_breakdown}
\resizebox{\textwidth}{!}{%
\begin{tabular}{l|ccccc}
\toprule
\textbf{Reasoning Type} & \textbf{Quant. Calc.} & \textbf{Pattern Ident.} & \textbf{Comp. Anal.} &  \textbf{Logic. Ded.} & \textbf{Comp. Synth.} \\ 
\midrule
&\multicolumn{5}{c}{\textit{Proprietary Models}} \\  \midrule
Gemini-3.1 Pro & \textbf{51.2}  & \textbf{68.5} / 3.6 & \textbf{73.8} / 3.7 & \textbf{87.3} / 4.3 & \textbf{77.3} / 4.2 \\
GPT-5.4 & 47.3  & 64.4 / \textbf{3.9} & 69.5 / \textbf{3.9} & 83.5 / \textbf{4.5} & 74.8 / \textbf{4.6} \\
Doubao-1.6 Vision & 40.0  & 60.7 / 3.6 & 64.7 / 3.7 & 78.5 / 4.3 & 68.0 / 4.2 \\
GPT-5 & 38.6  & 61.1 / 3.7 & 60.7 / 3.8 & 80.3 / 4.5 & 66.7 / 4.4 \\
Gemini-2.5 Pro & 29.9  & 56.6 / 3.7 & 58.3 / 3.4 & 79.2 / 4.3 & 72.2 / 3.9 \\
Gemini-2.5 Flash & 24.2  & 52.4 / 3.3 & 51.2 / 3.1 & 73.3 / 4.3 & 60.3 / 4.0 \\
Qwen3-Max & 8.3  & 40.3 / 3.1 & 35.6 / 2.7 & 64.7 / 3.9 & 52.0 / 3.5 \\
Grok-4.1 Fast & 7.5  & 39.9 / 2.4 & 37.1 / 2.1 & 68.2 / 3.7 & 48.1 / 3.3 \\ 
\midrule
&\multicolumn{5}{c}{\textit{Open-Source Models}} \\  \midrule
Qwen3-VL-235B & 41.1  & 60.2 / 3.5 & 64.4 / 3.3 & 77.2 / 4.4 & 70.7 / 4.1 \\
Qwen3-VL-30B & 32.2  & 53.4 / 3.4 & 57.4 / 2.7 & 72.1 / 4.0 & 54.7 / 3.8 \\
Qwen2.5-VL-72B & 14.3  & 49.1 / 2.8 & 42.5 / 2.5 & 70.0 / 3.7 & 52.0 / 3.2 \\ 
\bottomrule
\end{tabular}%
}
\end{table*}

\subsection{Multi-Dimensional Analysis}
\label{sec:fine_grained}

\noindent\textbf{Main Results.}
Tab.~\ref{tab:main_results} summarizes the performance of the eleven evaluated LMMs against the expert baseline.
Results indicate that the proprietary Gemini-3.1 Pro (64.12) and the open-weights Qwen3-VL-235B (57.73) achieve the highest overall scores among the evaluated AI models.
Notably, Qwen3-VL-235B rivals commercial models, attaining high accuracy in \textit{single\_choice} (75.25\%) and \textit{numerical\_precise} (52.89\%).
However, with no model surpassing an overall score of 65, a substantial gap of approximately 15 points remains compared to the \textbf{Human Expert} baseline (79.41), demonstrating that \textsc{FinDocMRE} presents a formidable challenge.

A marked performance disparity exists between visual perception and quantitative reasoning.
While models perform well on single-choice questions, accuracy declines on \textit{numerical\_approximate} tasks, where the top-performing Gemini-3.1 Pro reaches only 37.82\%.
Since \textit{numerical\_approximate} tasks require estimating values directly from figures, this suggests that current LMMs lack precision in trend estimation and arithmetic under visual ambiguity.
However, in \textit{open\_ended} scenarios, GPT-5.4 leads with a semantic score of 4.41/5.00, even outperforming the human experts (4.25), indicating strong capabilities in synthesizing financial narratives despite calculation limitations.

\noindent\textbf{Performance by Financial Domain.}
Tab.~\ref{tab:domain_breakdown_part1} and~\ref{tab:domain_breakdown_part2} illustrate performance variations across 12 financial domains.
A divergence exists between calculation and synthesis: while Gemini-3.1 Pro leads in quantitative domains such as Valuation (75.5\%) and M\&A (66.3\%), GPT-5.4 outperforms in qualitative reasoning, achieving the highest open-ended scores in 11 categories.
This suggests distinct optimization strategies: certain models emphasize visual grounding for extraction, while others focus on semantic synthesis for narrative construction.
Furthermore, a performance gap separates standardized "micro" domains (\emph{e.g.}, Risk Management, ~68\% accuracy) from ``macro'' domains (\emph{e.g.}, Fund Performance, <60\%), implying that LMMs face greater challenges interpreting market trendlines compared to structured formats.
Notably, Gemini-3.1 Pro shows strong generalization in Fintech (66.1\%), successfully processing the conceptual diagrams.

\noindent\textbf{Performance by Reasoning Type.}
Tab.~\ref{tab:reasoning_breakdown} details model performance across five reasoning categories, revealing distinct performance profiles.
While Gemini-3.1 Pro and GPT-5.4 lead in \textit{Quantitative Calculation} ($51.2\%$ and $47.3\%$ respectively), surpassing GPT-5 ($38.6\%$), this trend reverses in tasks requiring abstract inference.
In \textit{Logical Deduction}, Gemini-3.1 Pro achieves the highest accuracy ($87.3\%$) , while GPT-5.4 achieves the highest semantic score ($4.5$), indicating strong logical reasoning capabilities.
In \textit{Comprehensive Synthesis}, results diverge: Gemini-3.1 Pro attains the highest objective accuracy ($77.3\%$), while GPT-5.4 leads in semantic quality ($4.6/5.0\%$), suggesting that while some models excel at verification, GPT-5.4 is effective at constructing coherent financial narratives.

\noindent\textbf{Impact of Model Scale and Evolution.}
We use the Qwen series to distinguish the effects of parameter scale from architectural evolution.
Within the v3 generation, scaling laws persist; the 235B model ($57.73$) exceeds the 30B variant ($49.81$) by approximately 8 points, with the gap widening in calculation-heavy tasks like \textit{Numerical Precise} ($52.89\%$ vs $41.16\%$), suggesting larger capacity supports symbol manipulation.
However, inter-generational comparison shows that the Qwen3-VL-30B outperforms the previous Qwen2.5-VL-72B ($36.14$) by $13.67$ points.
This gain is evident in fine-grained reasoning, where Qwen3-30B nearly triples the accuracy of Qwen2.5-72B in quantitative calculation ($41.16\%$ vs $14.10\%$), indicating that algorithmic optimizations and multimodal data alignment are more critical for financial reasoning than raw parameter count. 

\begin{figure}[t]
    \centering
    \begin{minipage}[c]{0.49\linewidth}
        \centering
        \includegraphics[width=\linewidth]{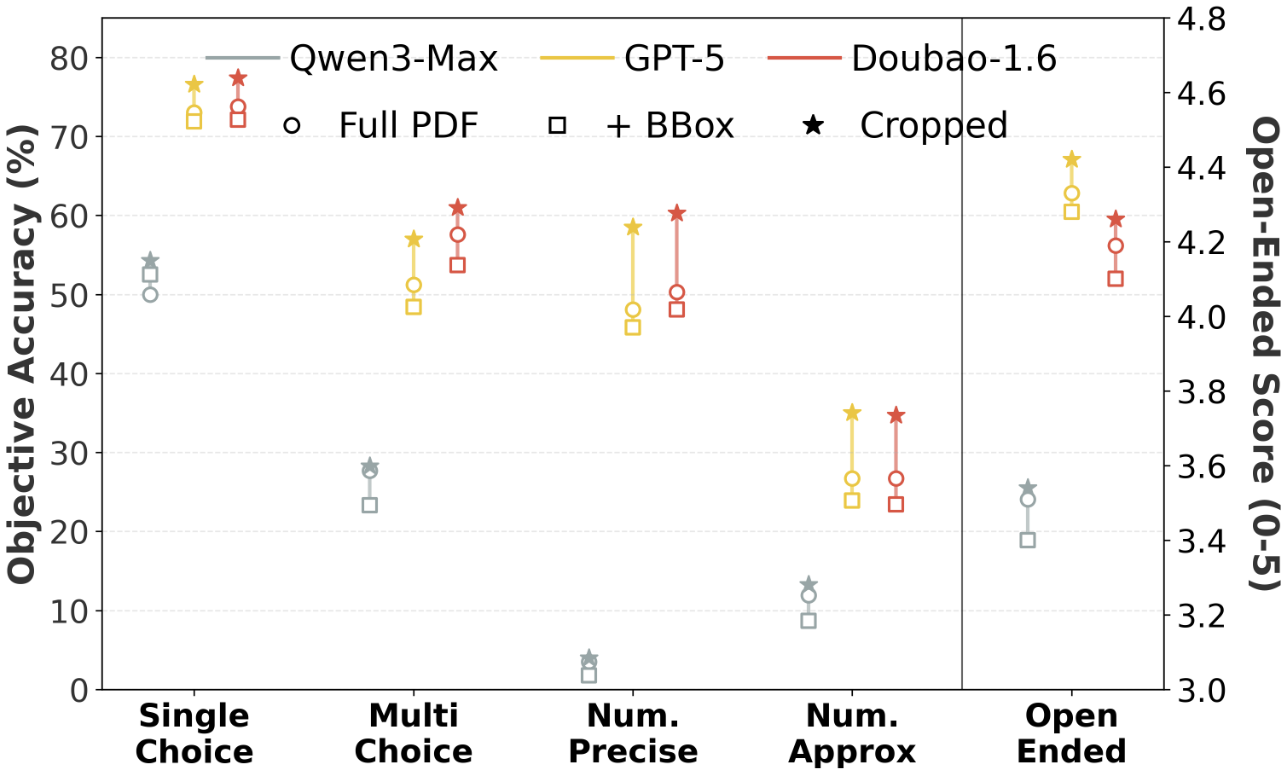} 
        \captionof{figure}{Impact of visual context on objective accuracy. While all models benefit from Cropped Images, Bounding Box annotations degrade the performance of advanced models, due to occlusion or visual noise.}
        \label{fig:ablation_context}
    \end{minipage}
    \hfill 
    \begin{minipage}[c]{0.49\linewidth}
        \centering
        \captionof{table}{Performance stratified by image count per sample.
        \textbf{S}, \textbf{M}, and \textbf{L} denote 1, 2, and $\ge$3 images.
        Cell format: \textbf{Accuracy (\%)} / \textbf{Score (0-5)}.}
        \label{tab:size_breakdown}   
        \setlength{\tabcolsep}{3.5pt} 
        \resizebox{\linewidth}{!}{%
        \begin{tabular}{l|ccc}
            \toprule
            \textbf{Image Quantity} & \textbf{S (1 img)} & \textbf{M (2 imgs)} & \textbf{L ($\ge$3 imgs)} \\ 
            \midrule
            &\multicolumn{3}{c}{\textit{Proprietary Models}} \\ 
            \midrule
            Gemini-3.1 Pro & \textbf{61.30} / 4.09 & \textbf{57.59} / 4.10 & \textbf{57.20} / 4.16 \\
            GPT-5.4 & 57.32 / \textbf{4.29} & 55.40 / \textbf{4.42} & 54.75 / \textbf{4.45} \\
            Doubao-1.6 Vision & 54.42 / 4.12 & 53.05 / 4.18 & 51.89 / 4.23 \\
            GPT-5 & 53.51 / 4.21 & 50.67 / 4.35 & 47.50 / 4.37 \\
            Gemini-2.5 Pro & 46.85 / 4.06 & 46.21 / 4.02 & 45.86 / 3.74 \\
            Gemini-2.5 Flash & 41.59 / 3.80 & 39.59 / 3.99 & 36.87 / 3.98 \\
            Qwen3-Max & 25.21 / 3.49 & 26.66 / 3.61 & 24.00 / 3.44 \\
            Grok-4.1 Fast & 25.76 / 3.19 & 26.51 / 3.39 & 24.82 / 3.20 \\ 
            \midrule
            &\multicolumn{3}{c}{\textit{Open-Source Models}} \\ 
            \midrule
            Qwen3-VL-235B & 55.09 / 4.14 & 52.89 / 4.12 & 51.99 / 4.04 \\
            Qwen3-VL-30B & 47.72 / 3.72 & 44.82 / 3.73 & 41.78 / 3.76 \\
            Qwen2.5-VL-72B & 32.76 / 3.39 & 32.91 / 3.29 & 28.60 / 3.04 \\ 
            \bottomrule
        \end{tabular}%
        }
    \end{minipage}
\end{figure}

\noindent\textbf{Impact of Visual Complexity.}
We evaluate cross-chart reasoning by categorizing related image samples into Single~(S), Dual~(M), and Multiple~(L, $\ge$3) groups.
As shown in Tab.~\ref{tab:size_breakdown}, performance in objective tasks declines as visual dependency increases.
For instance, Gemini-3.1 Pro's accuracy drops from 61.30\% (S) to 57.20\% (L), indicating difficulty in aggregating information across distributed visual inputs.
In contrast, open-ended scores remain stable or improve in complex settings (\emph{e.g.}, GPT-5.4 reaches 4.45 in L).
This suggests that while models struggle with precise grounding from multiple charts, they effectively synthesize comprehensive narratives from large visual contexts.

\noindent\textbf{Performance by Language.}
Evaluating across languages (Tabs.~\ref{tab:english_results} and~\ref{tab:chinese_results} in Appendix~\ref{app:extra_experiment}) reveals a \textit{Language Alignment Bias}.
Gemini-3.1 Pro leads the English subset (69.56\%), while GPT-5.4 excels in Open-Ended tasks (4.58/5.0).
Conversely, the Chinese subset exhibits a \textit{Home Field Advantage} for native architectures: Doubao-1.6 Vision remains highly competitive (59.92\%), surpassing GPT-5.4 (58.24\%) and the open-weights Qwen3-VL-235B (57.65\%) due to superior visual grounding in character-dense charts.
Remarkably, despite weaker Chinese visual perception, GPT-5.4 retains the highest Open-Ended score across both subsets.
This indicates a \textit{decoupling of perception and reasoning}, allowing these models to generate coherent narratives from imperfectly grounded data.\textbf{More experiments are in Appendix~\ref{app:extra_experiment}.}

\subsection{Ablation Studies}
\label{sec:ablation}


\noindent\textbf{Impact of Visual Context Strategy.}
To evaluate the effect of visual context, we conduct an ablation study using three strategies:
1) \textbf{Full PDF} (Baseline), using complete document pages;
2) \textbf{PDF + BBox}, overlaying red bounding boxes to highlight relevant figures; 
3) \textbf{Cropped Image}, providing extracted figures at original resolution. As illustrated in Fig.~\ref{fig:ablation_context}, comparing the \textbf{Cropped Image} setting with the \textbf{Full PDF} baseline reveals a distinct performance gap.
Leading models like Doubao-1.6 Vision and GPT-5 achieve their highest accuracy in the Cropped setting (\emph{e.g.}, Doubao-1.6 Vision shows marked gains~(over 9\%) in Numerical Precise), indicating that locating dispersed visual evidence in multi-page documents remains a major challenge.
However, the \textbf{PDF + BBox} strategy yields unexpected results.
Instead of improving performance, bounding box hints often degrade accuracy for high-capacity models (\emph{e.g.}, GPT-5 scores lower in Multi-Choice with BBox than with Full PDF).
We hypothesize that superimposed frames introduce visual interference or occlusion, disrupting character recognition and feature encoding rather than aiding navigation.


\noindent\textbf{Robustness to Prompt Variation.}
To verify the stability of model rankings against prompt sensitivity, we evaluate performance under prompt variations on a randomly sampled 10\% subset of objective questions. We compare the \textbf{Standard Prompt} used in our main experiments against a \textbf{Chain-of-Thought (CoT) Prompt}, created by appending \textit{``Please think step-by-step for the final answer''} to the original instructions in order to elicit explicit logical reasoning paths.

\begin{table}[h]
\centering
\caption{Stability under prompt variation. Accuracy on a 10\% randomly sampled objective subset.}
\setlength{\tabcolsep}{6pt} 
\begin{tabular}{lccc}
  \toprule
  \textbf{Model} & \textbf{Standard Acc.} & \textbf{CoT Acc.} & \textbf{$\Delta$} \\
  \midrule
  \textbf{Qwen3-VL-235B}  & 54.05\% & 54.27\% & +0.22\% \\
  \textbf{Doubao-1.6}     & 53.71\% & 53.69\% & -0.02\% \\
  \textbf{GPT-5}          & 51.96\% & 52.11\% & +0.15\% \\
  \textbf{Gemini-2.5-Pro} & 46.54\% & 46.41\% & -0.13\% \\
  \bottomrule
\end{tabular}%
\label{tab:prompt_stability}
\end{table}
As detailed in Table~\ref{tab:prompt_stability}, performance fluctuations are marginal across all tested models (all absolute $\Delta < 1\%$). More importantly, the relative model rankings remain entirely unchanged. This confirms that the evaluated performance accurately reflects the models' intrinsic document understanding capabilities, demonstrating robust stability against prompt phrasing variations.

\subsection{Case Study}
\begin{figure*}[h] 
    \centering
    \includegraphics[width=\textwidth]{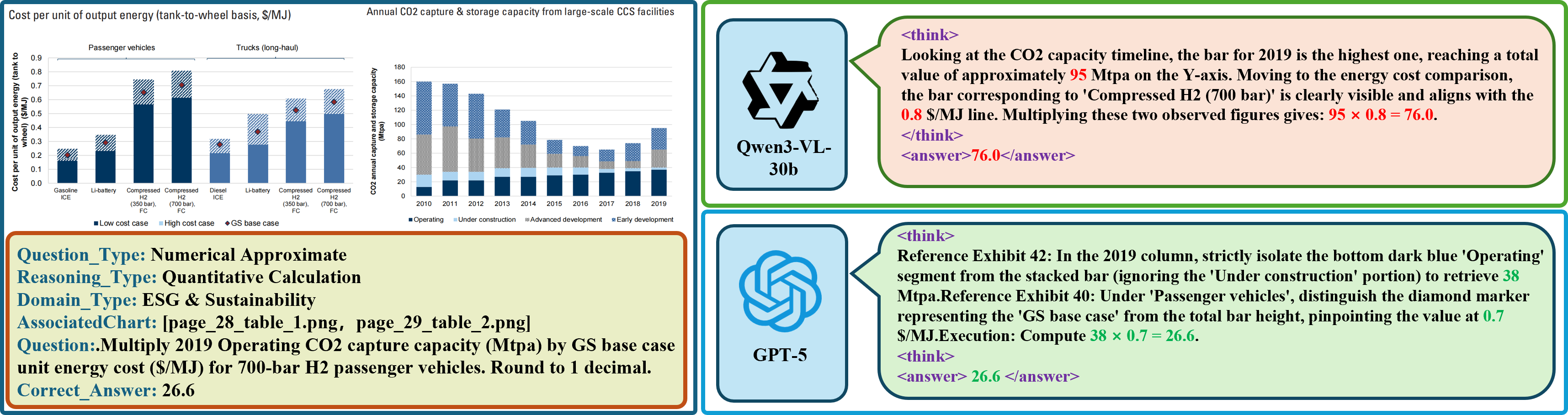} 
    \caption{
        Comparison of GPT-5 and Qwen3-VL-30B on a cross-page calculation task.
    }
    \label{fig:case_study}
\end{figure*}

Fig.~\ref{fig:case_study} illustrates a cross-page multiplication task testing fine-grained visual grounding, requiring the separation of stacked data series and legend-based markers.
Although Qwen3-VL-30b locates the target charts, its reasoning trace indicates susceptibility to \textbf{visual saliency bias}.
Despite the prompt requesting ``Operating'' capacity and ``GS base case'' cost, the model still selects the total bar height ($\sim$95) rather than the ``Operating'' segment ($\sim$38) and aligns with the bar's ceiling ($0.8$) instead of the diamond marker ($0.7$).
Therefore, disregarding legend constraints results in a calculation error ($95 \times 0.8 = 76.0$) relative to the ground truth ($26.6$).
In contrast, GPT-5 exhibits effective \textbf{legend-to-pixel alignment}, filtering visual noise (\emph{e.g.}, the ``Under construction'' stack) to locate data points defined by semantic constraints. \textbf{More case studies are in Appendix~\ref{app:extra_case_study}.}

\section{Conclusion}
\label{sec:conclusion}

In this paper, we introduce \textsc{FinDocMRE}, a benchmark designed to advance multimodal financial reasoning from isolated charts to document-level multimodal contexts.
Using a pipeline that combines visual-centric generation with expert verification, we compiled 12,207 samples from 2,878 financial reports to identify limitations in current SOTA models.
Extensive experiments across all eleven evaluated LMMs reveal a ``Analyst-Calculator Dichotomy'': while models show strong semantic synthesis, they lack the numerical precision and cross-page visual grounding required for professional analysis.
Ablation studies identify an ``visual grounding bottleneck'', confirming that the primary challenge lies in aggregating fragmented evidence from complex reports.
Based on these findings, future research should shift from end-to-end generation to agentic architectures capable of long-context grounding and tool-assisted calculation, fostering trustworthy financial AI.

\section{Limitations}
\label{sec:limitations}
Despite the scale of \textsc{FinDocMRE}, we acknowledge two primary limitations. 
First, our evaluation is restricted to zero-shot settings. Since each sample requires including an entire PDF document and its question and answer, providing few-shot exemplars would produce excessive prompt lengths that exceed most models' context window constraints. 
Second, our single-turn QA formulation simplifies complex, iterative professional financial workflows. Our benchmark currently focuses on foundational reasoning units, reserving multi-turn, agentic evaluations for subsequent research. 

{
\small
\bibliographystyle{neurips_2026}
\bibliography{neurips_2026}
}

\newpage
\appendix
Sec.~\ref{app:benchmark_construction} elaborates on the construction of \textsc{FinDocMRE}, documenting prompts and settings.
Sec.~\ref{app:evaluation_protocol} outlines the evaluation protocol, including configuration and scoring prompts.
Sec.~\ref{app:extra_experiment} presents additional analysis regarding language, visual resolution, and document length.
Sec.~\ref{app:extra_case_study} examines two additional cases illustrating visual bias and domain-specific capabilities of LMMs.

\section{\textsc{FinDocMRE} Construction}\label{app:benchmark_construction}

\subsection{Data Preparation}
To construct a benchmark reflecting real-world financial analysis, we collect a corpus of 2,878 financial PDFs from official sources.
We use a chart-centric extraction framework based on PyMuPDF and pdfplumber to parse both textual and embedded visual content.
Extracted text is serialized into JSON format to preserve document structure, while embedded charts are archived by source document.
Each image is assigned a filename reflecting its page location and extraction sequence (\emph{e.g.}, page\_1\_image\_1.png) as the unique chart\_id.

The raw extraction output contains noise, including company logos, icons, and fragmented artifacts such as decorative headers or dividers.
Furthermore, the initial extraction includes incomplete figures and text blocks incorrectly identified as images.
To isolate analytically valuable charts, we apply a Joint-Rule Filtering Mechanism:
\begin{enumerate}
    \item We apply geometric constraints to eliminate layout artifacts, discarding images with low resolutions or extreme aspect ratios.
    \item We integrate image similarity computation with OCR to remove repetitive non-data elements. This hybrid approach enables the deduplication of visual content and the removal of uninformative icons or corporate logos.
    \item We perform page-level textual indexing verification. By scanning the text of corresponding pages, we retain only images explicitly referenced by narrative markers (\emph{e.g.}, ``Figure x''), ensuring that all ``Cleaned Figures`` are linked to the document context.
\end{enumerate}

\subsection{Visual-Centric Generation}
To reduce textual bias and ground reasoning in visual evidence, we use a visual-centric generation strategy with \textsc{[Gemini-2.5-Pro]}. 
Instead of full PDF pages, we input the sequence of ``Cleaned Figures`` from the data preparation phase, excluding surrounding textual context.
Guided by a structured prompt, the model generates a standardized metadata object for each sample, containing eight specific fields: \textit{Question}, \textit{Options} (empty for non-choice tasks), \textit{Answer}, \textit{Question Type}, \textit{Domain Type}, \textit{Reasoning Type}, \textit{Associated Chart} (linking to specific chart IDs), and \textit{Reasoning Trace}.
This strict schema provides broad coverage of financial reasoning dimensions.
The prompt used for this generation process is presented in Tab.~\ref{tab:prompt_full}.

\subsection{Expert Verification}
\label{app:human_guidelines}

To reduce hallucination risks associated with visual-centric generation and ensure data quality, we employ human experts to verify each sample.
Three senior financial researchers with over three years of professional experience in securities firms verify each document and generated sample.
All the experts are Master in Economics or Finance from QS Top-100 universities, who have received compensation exceeding local minimum wage~(\$50/hour).
Guided by the annotation protocols in Tab.~\ref{tab:human_guidelines}, experts audit samples by cross-referencing generated reasoning traces with source charts.

\subsection{Reasoning Type and Financial Industries}\label{app:reasoning_and_domains}

To assess the viability of LMMs as professional financial agents, we established a multi-dimensional taxonomy evaluating \textbf{Cognitive Depth}~(Reasoning Types) and \textbf{Knowledge Breadth}~(Financial Domains).
This framework differentiates elementary visual reasoning from analytical logic.

\subsubsection{Reasoning Types}
Drawing from Bloom's Taxonomy~\cite{anderson2001taxonomy}, we classify reasoning skills into five levels, ranging from reasoning to synthesis. 
This hierarchy enables the identification of specific deficiencies in multimodal models.
\begin{itemize}[leftmargin=*]
    \item \textbf{Quantitative Calculation (Computation):} Fundamental to financial analysis, this category evaluates the capacity for precise multi-step arithmetic (\emph{e.g.}, CAGR, margins) using extracted visual data, verifying computational reliability.
    \item \textbf{Pattern Identification (Perception):} This assesses the recognition of visual trends, volatility, and anomalies (\emph{e.g.}, detecting a ``double top'' or sudden revenue drop), a capability exclusive to the visual modality.
    \item \textbf{Comparative Analysis (Relation):} Reflecting the benchmarking workflow, this requires aligning and contrasting data points across entities (\emph{e.g.}, ``Company A vs. Company B'') or time periods (YoY growth) to test relational reasoning.
    \item \textbf{Logical Deduction (Inference):} This evaluates the application of implicit financial rules (\emph{e.g.}, ``If assets increased while liabilities remained flat, equity must have increased''), evaluating adherence to domain logic.
    \item \textbf{Comprehensive Synthesis (Integration):} This requires aggregating fragmented information from heterogeneous charts (\emph{e.g.}, combining macro GDP with micro revenue) to form a unified assessment, similar to investment memo creation.
\end{itemize}

\subsubsection{Financial Domains}
To guarantee benchmark diversity and prevent overfitting to specific document types (\emph{e.g.}, Annual Reports), we select 12 domains covering the ``Micro-Meso-Macro'' spectrum of the financial ecosystem.

\begin{enumerate}[leftmargin=*]
    \item \textbf{Micro-Level (Corporate Finance):} Concentrating on firm-specific analysis.
    \begin{itemize}
        \item \textit{Financial Statement Analysis}: Fundamental accounting evaluation~(\emph{e.g.}, Income/Balance Sheets).
        \item \textit{Valuation}: Asset pricing methodologies~(DCF, P/E ratios).
        \item \textit{M\&A and Capital Structure}: Corporate transactions and capital composition~(\emph{e.g.}, Debt/Equity, Mergers).
        \item \textit{Business Segments}: Disaggregated analysis of revenue sources by region or product.
        \item \textit{Risk Management}: Assessment of exposure and stability (\emph{e.g.}, VaR, Liquidity).
    \end{itemize}
    
    \item \textbf{Meso-Level (Market \& Investment):} Analysis of portfolios and industrial sectors.
    \begin{itemize}
        \item \textit{Industry \& Competitive Analysis}: Evaluation of market share and strategic positioning (\emph{e.g.}, SWOT).
        \item \textit{Portfolio Management}: Asset allocation and diversification strategies.
        \item \textit{Fund Performance}: Attribution analysis (\emph{e.g.}, Alpha, Beta) of investment funds.
        \item \textit{Fintech \& Innovation}: Emerging trends in digital finance (\emph{e.g.}, Blockchain, AI).
    \end{itemize}
    
    \item \textbf{Macro-Level (Global Economy):} Examination of the global economic environment.
    \begin{itemize}
        \item \textit{Macroeconomic Indicators}: Primary economic drivers (\emph{e.g.}, GDP, Inflation, Rates).
        \item \textit{Macro Markets}: Commodity trends (\emph{e.g.}, Gold/Oil) and foreign exchange markets.
        \item \textit{ESG \& Sustainability}: Non-financial metrics regarding sustainability and governance (\emph{e.g.}, Carbon, Compliance).
    \end{itemize}
\end{enumerate}

This domain diversity enables \textsc{FinDocMRE} to evaluate LMMs beyond basic OCR capabilities, testing the interpretation of specialized semantics across the financial sector.

\newpage
\noindent\rule{\textwidth}{1.5pt} 
\begin{center}
    \textbf{\large System Prompt for Financial Reasoning Generation}
\end{center}
\noindent\rule{\textwidth}{0.5pt} 


\textbf{\# Role} \\
You are a senior financial analyst and an expert in designing challenging logical reasoning problems. Your mission is to create a challenging benchmark capable of effectively distinguishing the multimodal capabilities of current and future State-of-the-Art (SOTA) models.

\textbf{\# Core Philosophy}

\textbf{1. Form Serves Function:} Strictly adhere to question formats: \texttt{single\_choice}, \texttt{multiple\_choice}, \texttt{numerical\_precise} (Calculations must use \textbf{explicit values} found directly on charts/tables without ambiguity), \texttt{numerical\_approximate} (Reasoning requires \textbf{visual estimation} \emph{e.g.}, reading axis height where precise labels are absent), and \texttt{open\_ended} (Pure text answer; strictly forbidden from containing \texttt{chart\_id} or filenames).

\textbf{2. Evaluate Reasoning, Not Memorization:} The primary objective is evaluating deep analytical, reasoning, and synthesis skills.

\textbf{3. Blind Stem Principle:} The \texttt{stem} is strictly forbidden from mentioning \texttt{chart\_id} so users emulate real-world blind queries.

\textbf{4. Promote Comprehensive Analysis:} Encourage the design of complex questions that require integrating partial information from multiple distinct charts.

\textbf{5. Information Silo Principle:} All charts must be treated as originating from a fictional, non-public context; do not use external knowledge.

\textbf{6. Abstract Time Principle:} Avoid real dates. Use relative years (\emph{e.g.}, 'Year 1', 'Year 2'). If this conflicts with 'Event Anchoring', the latter takes precedence.

\textbf{7. Quantitative Anchor Principle:} Answers must be uniquely determined by specific information in the charts, avoiding ambiguous estimation scenarios.

\textbf{8. Event Anchoring Principle:} Prioritize using specific events (\emph{e.g.}, "when revenue peaked") to lock time points across multiple charts.

\textbf{9. Context-Free Stem Principle:} The stem must be clear and unambiguous, ensuring solvability whether the input is the full PDF or filtered images.

\textbf{\# Classification Tags} \\
When generating each question object, you must also add the following two classification tags. The definitions are strict:

\textbf{1. \texttt{5 reasoning\_type} (Reasoning Type):} [Must choose one]

$\bullet$ \textbf{Quantitative Calculation}: The core task is to perform multi-step arithmetic or algebraic operations to arrive at a precise or approximate numerical value. \\
$\bullet$ \textbf{Pattern Identification}: The core task is to identify and locate a specific data point, trend, pattern, or anomaly (\emph{e.g.}, 'peak', 'fastest growth'). \\
$\bullet$ \textbf{Comparative Analysis}: The core task is to compare two or more entities or data points to determine their relationship (\emph{e.g.}, 'which is higher', 'are they correlated'). \\
$\bullet$ \textbf{Logical Deduction}: The core task is to apply an implicit rule, constraint, or financial identity to deduce a necessary conclusion (\emph{e.g.}, '...must be true?'). \\
$\bullet$ \textbf{Comprehensive Synthesis}: The core task is to integrate information from multiple heterogeneous charts to form a high-level, coherent summary, judgment, or evaluation.

\textbf{2. \texttt{12 domain} (Domain Tag):} [Must choose one]

$\bullet$ \textbf{Financial Statement Analysis}: Focuses on items and their ratios from financial statements like the Income Statement, Balance Sheet, and Cash Flow Statement. \\
$\bullet$ \textbf{Valuation}: Focuses on calculating or comparing a company's value (\emph{e.g.}, DCF, P/E, EV/EBITDA, comparable company analysis, etc.). \\
$\bullet$ \textbf{M\&A and Capital Structure}: Focuses on mergers, acquisitions, divestitures, debt, equity, leverage, etc. \\
$\bullet$ \textbf{Industry \& Competitive Analysis}: Focuses on market share, industry trends, competitive landscape, SWOT analysis, Porter's Five Forces, etc. \\
$\bullet$ \textbf{Business Segments}: Focuses only on unique metrics or drivers for specific company business segments (\emph{e.g.}, by product, by region), etc. \\
$\bullet$ \textbf{Portfolio Management}: Focuses on asset allocation, portfolio composition, diversification, benchmark comparison, etc. \\
$\bullet$ \textbf{Fund Performance \& Attribution}: Focuses on Alpha, Beta, Sharpe Ratio, tracking error, performance attribution, etc. \\
$\bullet$ \textbf{Risk Management}: Focuses on VaR (Value at Risk), credit risk, liquidity risk, operational risk, stress testing, etc. \\
$\bullet$ \textbf{Macroeconomic Indicators}: Focuses on GDP, inflation, interest rates, employment, consumer confidence indices, etc. \\
$\bullet$ \textbf{Macro Markets}: Focuses on the market dynamics of commodities (\emph{e.g.}, oil, natural gas, gold, etc.) and foreign exchange (\emph{e.g.}, currency pairs, exchange rate changes, etc.). \\
$\bullet$ \textbf{ESG \& Sustainability}: Focuses on environmental, social, and governance metrics, such as carbon emissions, diversity, governance structure, etc. \\
$\bullet$ \textbf{Fintech \& Innovation}: Focuses on digital payments, blockchain, AI in finance, RegTech, etc.

\textbf{\# Task Instructions} \\
Generate a JSON list that strictly adheres to the defined schema. \textbf{Special Rules}: 1. Return empty \texttt{[]} if review fails. 2. Balance categories.

\textbf{\# Step-by-Step Instructions}

\textbf{Step 1: In-depth Analysis}: Identify chart types, units, and axis meanings. Pay special attention to \textbf{real timestamps} and prepare to abstract them (\emph{e.g.}, converting dates to "Year 1"). Understand connections between charts.

\textbf{Step 2: Mine Scenarios}: Prioritize \textbf{"Event Anchoring"} (using events like "revenue peak" to lock time points across charts), visual geometric features (slopes, intersections), and scenarios requiring active filtering of key info from multiple charts.

\textbf{Step 3: Draft Questions}: Select format. \textbf{Hide calculation paths} (ask for the final result, not the steps). Ensure \textbf{no \texttt{chart\_id} in stems}. Design common-error distractors. Sanitize \texttt{open\_ended} answers (remove filenames). For numerical tasks, ensure multi-step calculation is required.

\textbf{Step 3.5: Adversarial Self-Review}: Perform a \textbf{"zero-knowledge test"}. Ask: "\textit{Based only on the text, can I deduce the answer using common sense without the chart?}" If \textbf{YES} (leaks knowledge), discard the question. If \textbf{NO}, it passes.

\textbf{Step 4: Generate JSON}: Add tags. Construct the \texttt{reasoning\_trace} to serve as a self-contained proof, explicitly stating which values are estimated vs. precise and explaining the derivation logic step-by-step.

\textbf{\# JSON Output Example (1 Shot)}

\begin{minipage}{\linewidth}
\begin{lstlisting}[
    basicstyle=\ttfamily\normalsize, % 使用打字机字体，正常大小
    columns=fullflexible,
    breaklines=true,      % 自动换行
    frame=single,         % 加框
    rulecolor=\color{black},
    language=json,
    keepspaces=true,
    aboveskip=0.5em,
    belowskip=0.5em,
    showstringspaces=false,
    morekeywords={question_id, source_chart_ids, question_format, reasoning_type, domain, question_content, stem, options, answer, reasoning_trace}
]
[
  {
    "question_id": "auto_generated_unique_id_1",
    "source_chart_ids": [ "page_6_image_1.png", "page_5_table_2.png" ],
    "question_format": "numerical_precise",
    "reasoning_type": "Quantitative Calculation",
    "domain": "Financial Statement Analysis",
    "question_content": {
      "stem": "Combining the sales trend chart and the cost structure table, in the year when sales revenue reached its historical peak, what was the company's raw material cost as a percentage of total cost? Present the answer as a percentage, rounded to one decimal place.",
      "options": {}
    },
    "answer": "42.5",
    "reasoning_trace": [
      "Self-Review: Passed. 'Sales revenue reached its historical peak' is an event requiring visual location. Cost percentage requires multi-step calculation.",
      "Step 1: Identify the 'Sales Trend Chart' (page_6_image_1.png) and 'Cost Structure Table' (page_5_table_2.png).",
      "Step 2: [Event Anchoring] In Sales Trend Chart, visually inspect the curve to find its highest point, determining the peak occurred in Year 5.",
      "Step 3: [Cross-Chart Query] Using Year 5 as anchor, query Cost Structure Table: Raw Material Cost = $3,400M, Total Cost = $8,000M.",
      "Step 4: [Precise Calculation] ($3,400M / $8,000M) * 100% = 42.5%."
    ]
  }
]
\end{lstlisting}
\end{minipage}

\noindent\rule{\textwidth}{1.5pt} 
\captionof{table}{The full system prompt used for financial reasoning question generation.}
\label{tab:prompt_full}
\begin{figure*}[h]
    \centering
    
     \includegraphics[width=0.8\columnwidth, height=0.5\columnwidth]
    {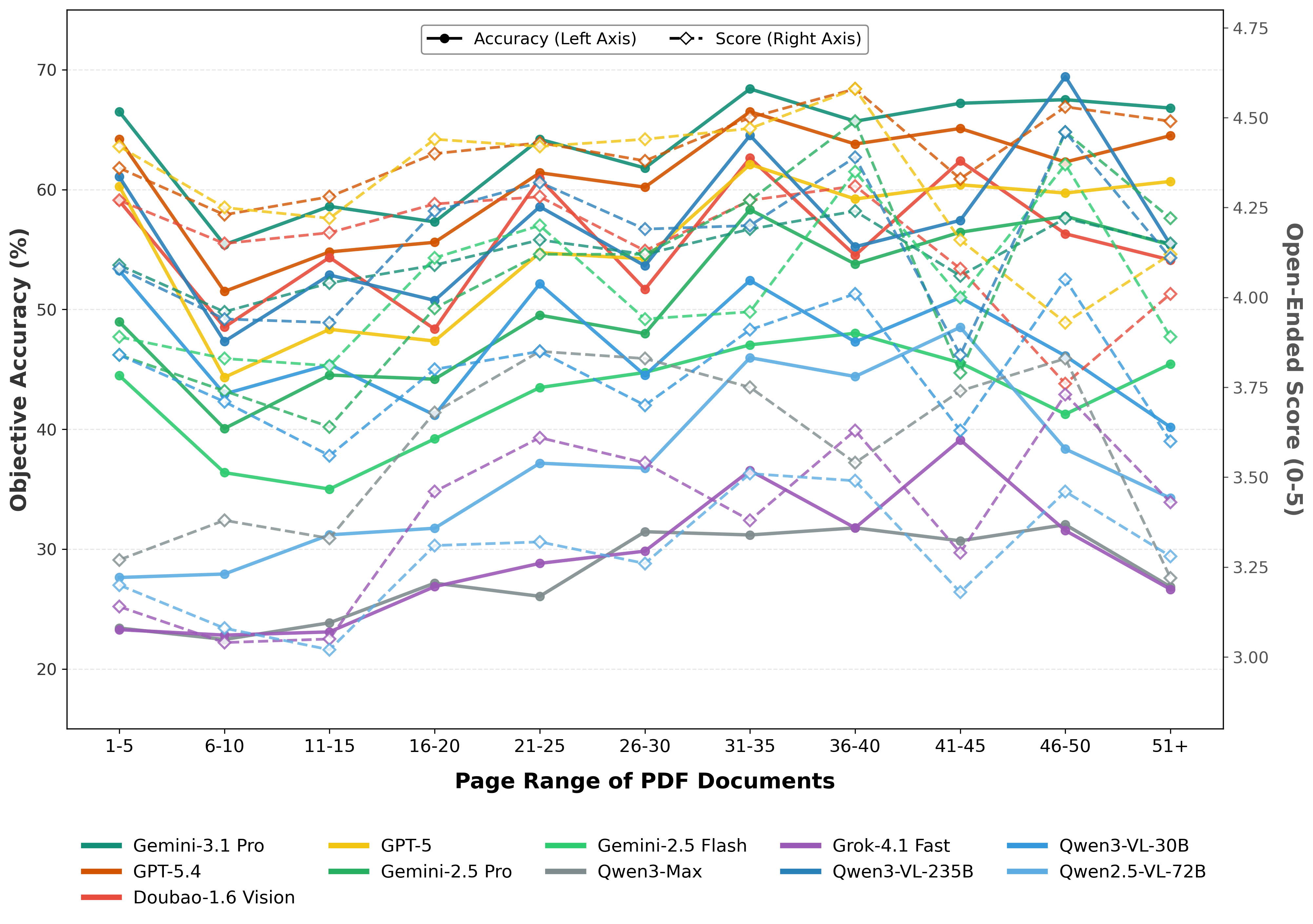}
    \caption{Impact of PDF Length}
    \label{fig:length_impact}
\end{figure*}

\clearpage
\raggedbottom   
\setlength{\parindent}{0pt}  
\setlength{\parskip}{0.6em}  


\noindent\rule{\textwidth}{1.5pt} 
\begin{center}
    \textbf{\large Expert Verification Guidelines}
\end{center}
\noindent\rule{\textwidth}{0.5pt} 


\textbf{Role:} You are a senior financial researcher acting as a data auditor. Your goal is to guarantee the benchmark's "Gold Standard" quality.

\textbf{Input Data Package:} \\
For each verification task, you will be presented with:
\textbf{(1) Source Context:} The set of extracted \texttt{Cleaned Charts} from the financial report.
\textbf{(2) Question Metadata:} The generated \texttt{Stem}, \texttt{Options} (for MCQs), and \texttt{Reference Answer}.
\textbf{(3) Logic Chain:} The step-by-step \texttt{Reasoning Trace} derived by the model.
\textbf{(4) Grounding:} The \texttt{Associated Chart IDs} indicating visual evidence sources.
\noindent\rule{\textwidth}{0.5pt}

\textbf{Objective:} Perform a binary validation (RETAIN vs. DISCARD). If a sample contains \textit{any} flaw, it must be discarded immediately.

\textbf{Phase 1: Visual Grounding Audit (Critical Check)}

$\bullet$ \textbf{Chart Alignment:} Verify that the \texttt{Associated Chart IDs} point \textit{exclusively} and \textit{correctly} to the images containing the evidence. If irrelevant charts are cited or key charts are missing $\to$ \textbf{DISCARD}.

$\bullet$ \textbf{Textual Isolation:} Ensure the question is solvable \textit{solely} using the provided visual charts. If the reasoning relies on external knowledge or non-visual PDF text (not present in the input charts) $\to$ \textbf{DISCARD}.

\textbf{Phase 2: Content \& Logic Audit}

$\bullet$ \textbf{Question Quality:} Is the phrasing clear and unambiguous? If the intent is vague or grammar is poor $\to$ \textbf{DISCARD}.

$\bullet$ \textbf{Factual Correctness:} Is the \texttt{Reference Answer} numerically precise and factually correct based strictly on the chart data? If any calculation error or data hallucination exists $\to$ \textbf{DISCARD}.

$\bullet$ \textbf{Option Quality (for MCQs):} Are the distractors plausible? If options are nonsensical or allow guessing without reasoning $\to$ \textbf{DISCARD}.

$\bullet$ \textbf{Trace Logic:} Does the \texttt{Reasoning Trace} provide a valid, step-by-step derivation? If the logic leaps or cites non-existent values $\to$ \textbf{DISCARD}.

\textbf{Phase 3: Final Decision (Unanimous Consensus Rule)}

$\bullet$ \textbf{\textcolor{red}{DISCARD (Action)}:} Select this if the sample triggers \textbf{ANY} violation from Phase 1 or 2.

$\bullet$ \textbf{\textcolor{green!60!black}{RETAIN (Action)}:} Select this \textit{only} if the sample is visually grounded, logically flawless, and explicitly verified by your professional judgment.

\noindent\rule{\textwidth}{1.5pt} 
\captionof{table}{Guidelines provided to human experts for verifying the quality of the generated benchmark data.}
\label{tab:human_guidelines}

\newpage
\section{Evaluation Protocol}\label{app:evaluation_protocol}

\subsection{Evaluation Details}
In this section, we detail the inference configuration, data input strategy, and scoring protocols.

\subsubsection{Inference and Input Settings}
We evaluate all models in a \textbf{Zero-Shot} setting using their official APIs with the inference prompt shown in Tab.~\ref{tab:eval_prompts1}. The detailed configurations:

\begin{itemize}[leftmargin=*]
    \item \textbf{Visual Input Strategy:} To standardize inputs, we convert all visual content (Full PDF, Bounding Box, or Cropped Charts) into Image Sequences. 
    Considering existing LMMs cannot directly process PDF files, we render visual contexts as high-resolution images.
    \item \textbf{Batching:} We query multiple questions associated with the same document in a single API session. 
    For documents with numerous questions (exceeding 10), we split them into separate batches.
    For each batch, we re-upload the associated images to maintain consistent visual context.
    \item \textbf{Hyper-parameter:} We use default inference parameters (\emph{e.g.}, \texttt{temperature}=0) to reflect baseline capabilities. 
    We fix the random seed at \texttt{1234} to support reproducibility.
    We enable ``thinking modes'' for models supporting reasoning traces.
    \item \textbf{Error Handling:} We implement a retry mechanism for standard API errors such as connection timeouts. For fatal errors (\emph{e.g.}, input limits or safety filters), we stop retries and record the answer as \texttt{null}~(treated as incorrect).
\end{itemize}

\subsubsection{Scoring Protocol}
Given that models output answers directly, we use Regular Expressions for standardization. The scoring protocols are defined as follows:

\begin{itemize}[leftmargin=*]
    \item \textbf{Single Choice: Exact Letter Match.} We compare the selected option with the ground truth.
    \item \textbf{Multiple Choice: Strict Set Match.} The predicted letter set must be identical to the ground truth, without partial credit.
    \item \textbf{Numerical Precise: Exact Equality.} The calculation correspond exactly to the ground truth.
    \item \textbf{Numerical Approximate: Conditional Tolerance.} We apply a two-step validation:
    \begin{enumerate}[label=(\roman*), nosep, leftmargin=15pt]
        \item If the reference value $V_{ref} = 0$, the prediction is correct only if $V_{pred} = 0$.
        \item If $V_{ref} \neq 0$, the prediction is correct if it falls within a $\mathbf{5\%}$ relative error margin: $|V_{pred} - V_{ref}| / |V_{ref}| \leq 0.05$.
    \end{enumerate}
    \item \textbf{Open Ended: LLM-as-a-Judge.} Three distinct LMMs evaluate the response with reference~(prompt in Tab.~\ref{tab:eval_prompts2}), assigning an \textbf{integer score} from [0, 5].
\end{itemize}

\subsection{Evaluation Bias}\label{app:bias}

To address concerns regarding evaluator subjectivity, we conduct a detailed bias and alignment analysis. Notably, approximately 80\% of the tasks in FinDocMRE (\emph{e.g.}, Multiple-Choice, Numerical) rely on deterministic metrics immune to subjectivity. For the remaining 20\% open-ended questions, we apply a reference-guided LMM-as-a-Judge approach. 
To mitigate single-model bias, our protocol ensembles three distinct models (\texttt{Gemini-2.5-Flash}, \texttt{GPT-4o}, and \texttt{Qwen3-Max}) to independently score responses, taking their average as the final metric. 
To validate this approach and assess its alignment with human experts, we conducted a comparative analysis on a sampled subset of 250 open-ended questions ($\sim$10\% of the open-ended test set). We compared the individual LMM scores against a \textbf{Human Expert} baseline, derived from the averaged independent ratings of two professional financial researchers. The evaluation results across five representative target models are presented in Table~\ref{tab:bias_analysis}.

\begin{table}[h]
\centering
\caption{Bias and alignment analysis on a sampled subset (250 questions). Results are \textbf{Score / Rank}. Target models evaluated: GPT-5, Doubao-1.6 Vision (DB-1.6), Qwen3-vl-235B (QW-235B), Gemini-2.5-Pro (Gem-Pro), and Qwen3-vl-30B (QW-30B).}
\setlength{\tabcolsep}{6pt}
\begin{tabular}{lccccc}
  \toprule
  \textbf{Evaluator} & \textbf{GPT-5} & \textbf{DB-1.6} & \textbf{QW-235B} & \textbf{Gem-Pro} & \textbf{QW-30B} \\
  \midrule
  \textbf{Gemini-2.5-Flash} & 4.32 / 1 & 4.22 / 2 & 4.09 / 3 & 3.91 / 4 & 3.75 / 5 \\
  \textbf{GPT-4o}           & 4.34 / 1 & 4.23 / 2 & 4.10 / 3 & 3.92 / 4 & 3.76 / 5 \\
  \textbf{Qwen3-Max}        & 4.36 / 1 & 4.25 / 2 & 4.12 / 3 & 3.95 / 4 & 3.79 / 5 \\
  \midrule
  \textbf{Human Expert}     & 4.11 / 1 & 3.96 / 2 & 3.71 / 3 & 3.38 / 4 & 3.19 / 5 \\
  \bottomrule
\end{tabular}%

\label{tab:bias_analysis}
\end{table}

As shown in Table~\ref{tab:bias_analysis}, while LMM evaluators exhibit slight leniency in absolute scoring (typically $+0.2$ to $+0.6$ higher than human experts), their relative rankings maintain alignment with human judgments. Across all evaluated target models, the performance hierarchy remains identical (1st to 5th) with minimal variance among the automated judges. This demonstrates that despite minor scaling differences in absolute scores, the ensembled LMMs reliably discern response quality, confirming that our evaluation framework is robust, objective, and aligned with professional standards.

\section{Extra Experimental Results}\label{app:extra_experiment}

\subsection{Performance by Language}

\begin{table}[h!]
\centering
\caption{Evaluation results for the \textbf{English subset}. Abbreviations denote question types: Single-Choice (\textbf{Single}), Multiple-Choice (\textbf{Multi}), Numerical-Precise (\textbf{N(p)}), Numerical-Approximate (\textbf{N(a)}), and Open-Ended (\textbf{Open}). \textbf{Overall} score is the weighted average.}
\setlength{\tabcolsep}{6pt}

\begin{tabular}{l|ccccc|c}
\toprule
\textbf{Model} & \textbf{Single} & \textbf{Multi} & \textbf{N(p)} & \textbf{N(a)} & \textbf{Open} & \textbf{Overall} \\
\midrule
 & \multicolumn{6}{c}{\textit{Proprietary Models}} \\ \midrule
Gemini-3.1 Pro & 82.32 & 65.50 & 69.10 & 43.90 & 4.35 (87.0) & 69.56 \\
GPT-5.4 & 79.77 & 62.23 & 65.15 & 37.07 & 4.58 (91.6) & 67.16 \\
GPT-5 & 76.32 & 53.93 & 49.56 & 27.61 & 4.37 (87.4) & 58.96 \\
Doubao-1.6 Vision & 73.77 & 53.28 & 45.51 & 24.88 & 4.15 (83.0) & 56.09 \\
Gemini-2.5 Pro & 76.01 & 48.25 & 38.30 & 20.20 & 4.26 (85.2) & 53.59 \\
Gemini-2.5 Flash & 68.92 & 46.94 & 28.13 & 21.17 & 4.03 (80.6) & 49.15 \\
Qwen3-Max & 55.58 & 30.79 & 5.73 & 14.44 & 3.76 (75.2) & 36.35 \\
Grok-4.1 Fast & 62.28 & 27.73 & 6.12 & 11.71 & 3.65 (73.0) & 36.17 \\
\midrule
 & \multicolumn{6}{c}{\textit{Open-Source Models}} \\ \midrule
Qwen3-VL-235B & 76.83 & 47.60 & 50.15 & 27.22 & 4.29 (85.8) & 57.52 \\
Qwen3-VL-30B & 70.33 & 29.69 & 36.62 & 19.41 & 3.75 (75.0) & 46.21 \\
Qwen2.5-VL-72B & 66.50 & 36.46 & 21.42 & 15.02 & 3.49 (69.8) & 41.84 \\
\bottomrule
\end{tabular}%

\label{tab:english_results}
\end{table}

\begin{table}[t]
\centering
\caption{Evaluation results for the \textbf{Chinese subset}. Abbreviations are question types: Single-Choice (\textbf{Single}), Multiple-Choice (\textbf{Multi}), Numerical-Precise (\textbf{N(p)}), Numerical-Approximate (\textbf{N(a)}), and Open-Ended (\textbf{Open}). \textbf{Overall} score are the weighted average.}
\setlength{\tabcolsep}{6pt}
\begin{tabular}{l|ccccc|c}
\toprule
\textbf{Model} & \textbf{Single} & \textbf{Multi} & \textbf{N(p)} & \textbf{N(a)} & \textbf{Open} & \textbf{Overall} \\
\midrule
 & \multicolumn{6}{c}{\textit{Proprietary Models}} \\ \midrule
Gemini-3.1 Pro & 76.44 & 54.51 & 58.91 & 33.45 & 3.95 (79.0) & 60.46 \\
Doubao-1.6 Vision & 73.77 & 59.69 & 53.81 & 27.92 & 4.22 (84.4) & 59.92 \\
GPT-5.4 & 72.92 & 54.09 & 51.36 & 27.01 & 4.29 (85.8) & 58.24 \\
GPT-5 & 70.25 & 49.84 & 46.98 & 26.10 & 4.30 (86.0) & 55.83 \\
Gemini-2.5 Pro & 69.93 & 43.94 & 34.70 & 21.06 & 3.65 (73.0) & 48.53 \\
Gemini-2.5 Flash & 60.90 & 39.17 & 22.99 & 20.78 & 3.88 (77.6) & 44.29 \\
Qwen3-Max & 45.25 & 26.22 & 1.80 & 10.01 & 3.32 (66.4) & 29.94 \\
Grok-4.1 Fast & 47.01 & 21.04 & 2.01 & 8.61 & 2.98 (59.6) & 27.65 \\
\midrule
 & \multicolumn{6}{c}{\textit{Open-Source Models}} \\ \midrule
Qwen3-VL-235B & 73.93 & 55.03 & 54.89 & 25.61 & 3.94 (78.8) & 57.65 \\
Qwen3-VL-30B & 67.79 & 49.95 & 44.47 & 21.62 & 3.73 (74.6) & 51.69 \\
Qwen2.5-VL-72B & 53.53 & 25.39 & 8.76 & 14.00 & 2.98 (59.6) & 32.26 \\
\bottomrule
\end{tabular}%

\label{tab:chinese_results}
\end{table}





\subsection{Evaluation Prompts}
\label{app:eval_prompts}
The full evaluation prompts used in our experiments are provided in Table~\ref{tab:eval_prompts1} and Table~\ref{tab:eval_prompts2}.

\begin{figure}[t]
    \centering
    \includegraphics[width=0.7\columnwidth, height=0.4\columnwidth]
    {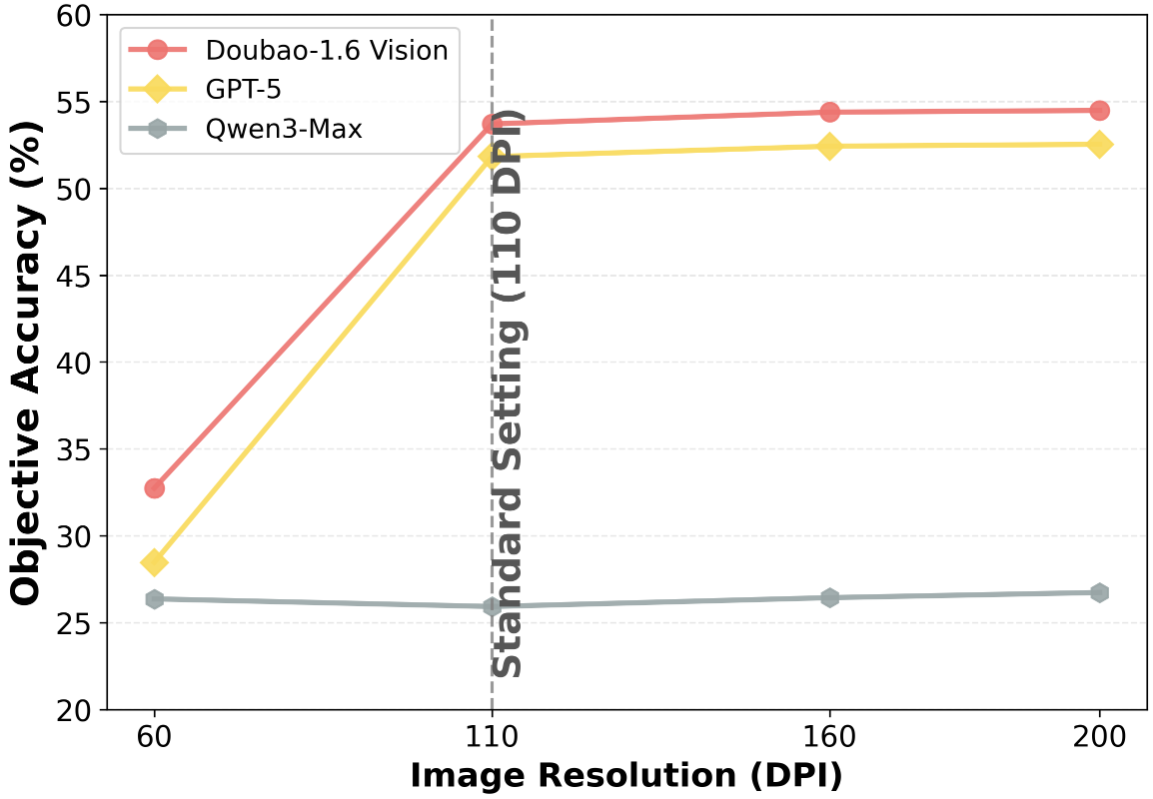}
    \caption{\textbf{Impact of Image Resolution (DPI).} 
    Advanced models (\emph{e.g.}, Doubao, GPT-5) show \textbf{minimal gains} beyond \textbf{110 DPI}, supporting the experimental setting. 
    Other models (\emph{e.g.}, Qwen3-Max) remain limited by reasoning capabilities regardless of resolution.}
    \label{fig:ablation_dpi}
\end{figure}

\subsection{Impact of Visual Resolution.}
We examine model sensitivity to image quality by varying resolution from 60 to 200 DPI.
As illustrated in Fig.~\ref{fig:ablation_dpi}, a performance saturation trend is observed.
For advanced models (Doubao-1.6 Vision, GPT-5), accuracy increases between 60 and 110 DPI as details become legible, stabilizing beyond 110 DPI.
This supports our selection of 110 DPI for main experiments, balancing visual clarity with token efficiency.
In contrast, lower-performing models (\emph{e.g.}, Qwen3-Max) show minimal gains, suggesting limitations in reasoning rather than visual perception.


\subsection{Impact of Document Length.}

Increasing source document length typically introduces visual noise, often causing performance degradation known as the ``Lost-in-the-Middle'' phenomenon.
However, as shown in Fig.~\ref{fig:length_impact}, results indicate a non-monotonic relationship between PDF page count and model performance, where accuracy fluctuates rather than declining consistently.
We attribute this behavior to three factors.
First, page count correlates with document standardization; longer documents are predominantly official Periodic Reports (\emph{e.g.}, Annual Reports) with professional typesetting and vector charts. In contrast, shorter documents show greater layout variance, increasing parsing difficulty.
Second, the Joint-Rule Filtering Mechanism stabilizes performance by reducing noise.
By removing decorative elements and text-only pages, the mechanism maintains high information density, preventing the processed visual context from expanding linearly with raw page count.
Finally, financial reasoning displays semantic clustering, where interdependent charts required for a query typically appear within the same thematic section (\emph{e.g.}, adjacent pages in the MD\&A chapter).
These characteristics enables models to process local information clusters without grounding disparate data across full document.

\section{Extra Case Study}\label{app:extra_case_study}
\subsection{Visual Bias in Comparative Analysis}
\begin{figure*}[h!] 
    \centering
    \includegraphics[width=\textwidth]{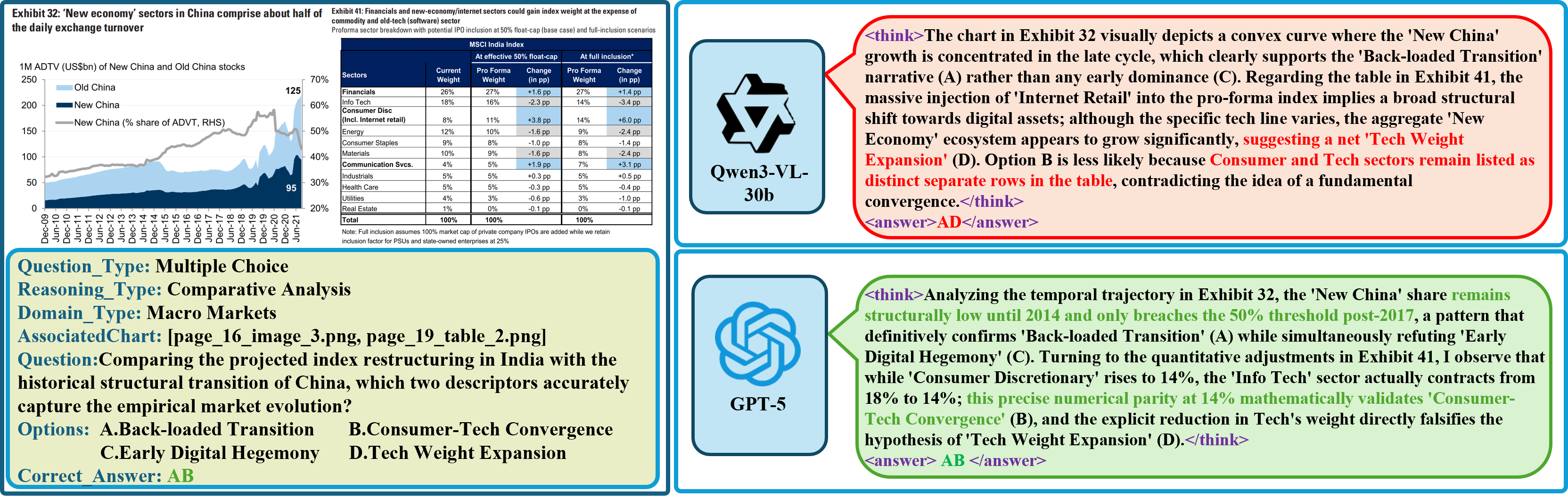} 
    \caption{Comparison of GPT-5 and Qwen3-VL-30B on Multiple Choice task.}
    \label{fig:case1}
\end{figure*}

In Fig.~\ref{fig:case1}, we present a task requiring cross-referencing charts to identify structural descriptors. 
GPT-5 correctly selects options A and B by calculating the exact parity (14\%) between sector weights, confirming ``Consumer-Tech Convergence.'' 
In contrast, Qwen3-VL-30B exhibits \textit{Visual Impression Bias}.
Although it correctly identifies the timeframe, it incorrectly selects ``Tech Weight Expansion'' (Option D), citing the visual addition of new rows as evidence of growth.
The model confuses \textit{visual layout expansion} with value increase, overlooking the numerical decrease from 18\% to 14\%.
This suggests that smaller models may prioritize visual heuristics (\emph{e.g.}, ``more rows implies growth'') over numerical extraction when visual signals conflict with quantitative data.

\subsection{Domain Cognition in Strategic Synthesis}
\begin{figure*}[h!] 
    \centering
    \includegraphics[width=\textwidth]{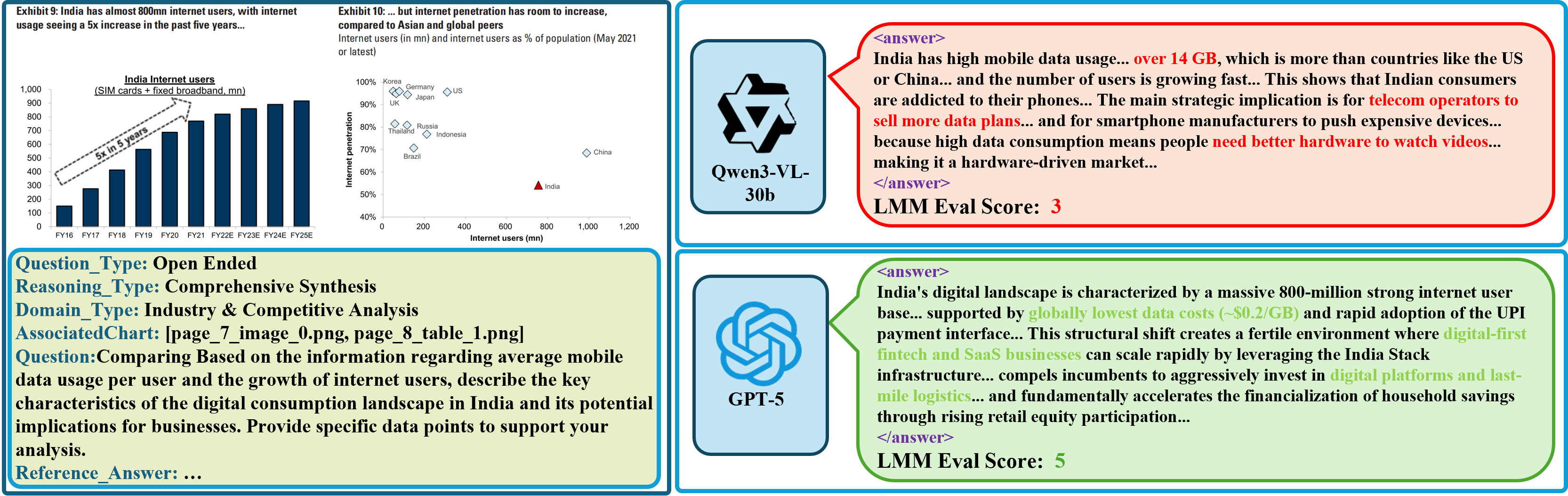} 
    \caption{Comparison of GPT-5 and Qwen3-VL-30B on Open Ended task.}
    \label{fig:case2}
\end{figure*}

In Fig.~\ref{fig:case2}, we evaluate strategic synthesis capabilities using an LLM-as-a-Judge scoring system. 
Qwen3-VL-30B scores 3; while factually accurate, it provides literal chart descriptions, limiting implications to generic consumer concepts like data plan sales.
In contrast, GPT-5 scores 5 by demonstrating domain alignment.
It infers second-order economic effects, connecting low data costs to ``digital-first fintech'' and the ``financialization of household savings'' concepts derived logically rather than visually.
This comparison illustrates that while visual perception is comparable, significant disparities remain in the ability to contextualize evidence within a professional framework.

\section{Broader Impacts}
Our work introduces FinDocMRE to advance document-level reasoning in financial AI, which can enhance the efficiency and transparency of financial information processing. However, we emphasize that this benchmark is for research purposes only and should not be used to generate automated investment advice or replace professional human judgment in financial decision-making.

\begin{table*}[h!]
\centering
\caption{Performance breakdown (Part 1/2): Business Segments to Industry Analysis. Metrics: \textbf{Accuracy (\%)} / \textbf{Score (0-5)}.}
\setlength{\tabcolsep}{12pt}

\resizebox{\linewidth}{!}{%
\begin{tabular}{l|cccccc}
\toprule
\textbf{Model} & \textbf{Bus. Seg.} & \textbf{ESG} & \textbf{Fin. Stat.} & \textbf{Fintech} & \textbf{Fund Perf.} & \textbf{Ind. Anal.} \\ 
\midrule
 & \multicolumn{6}{c}{\textit{Proprietary Models}} \\ \midrule
Gemini-3.1 Pro & 62.9 / 4.1 & 59.9 / 4.2 & 63.2 / 4.0 & 66.1 / 4.3 & 58.2 / 4.1 & 62.2 / 4.2 \\
GPT-5.4 & 60.6 / 4.3 & 58.3 / 4.4 & 58.2 / 4.4 & 63.9 / 4.4 & 55.3 / 4.2 & 59.8 / 4.5 \\
Doubao-1.6 Vision & 59.8 / 4.2 & 52.6 / 4.2 & 55.6 / 3.9 & 59.2 / 4.5 & 53.0 / 3.9 & 57.6 / 4.4 \\
GPT-5 & 56.8 / 4.3 & 56.0 / 4.3 & 50.9 / 4.2 & 61.2 / 4.4 & 44.9 / 3.8 & 55.9 / 4.5 \\
Gemini-2.5 Pro & 52.2 / 3.9 & 49.1 / 4.2 & 38.0 / 3.4 & 59.0 / 4.3 & 37.2 / 3.1 & 52.5 / 4.0 \\
Gemini-2.5 Flash & 45.8 / 3.8 & 38.3 / 4.0 & 28.8 / 3.6 & 51.8 / 4.4 & 26.2 / 3.2 & 44.0 / 4.2 \\
Qwen3-Max & 32.6 / 3.4 & 24.1 / 3.7 & 9.4 / 2.7 & 39.1 / 3.9 & 12.1 / 2.5 & 27.7 / 3.6 \\
Grok-4.1 Fast & 31.1 / 3.3 & 26.5 / 3.4 & 10.1 / 2.4 & 39.1 / 4.0 & 14.1 / 2.1 & 26.8 / 3.4 \\ 
\midrule
 & \multicolumn{6}{c}{\textit{Open-Source Models}} \\ \midrule
Qwen3-VL-235B & 58.9 / 3.9 & 55.9 / 4.2 & 51.8 / 3.7 & 62.0 / 4.5 & 52.8 / 3.7 & 57.9 / 4.3 \\
Qwen3-VL-30B & 51.5 / 3.6 & 42.7 / 3.8 & 47.8 / 3.6 & 55.0 / 4.1 & 45.7 / 3.3 & 48.1 / 4.1 \\
Qwen2.5-VL-72B & 38.6 / 3.4 & 33.3 / 3.3 & 15.4 / 2.7 & 47.5 / 3.6 & 19.1 / 2.7 & 37.0 / 3.3 \\ 
\bottomrule
\end{tabular}%
}

\label{tab:domain_breakdown_part1}
\end{table*}

\begin{table*}[h!]
\centering
\caption{Performance breakdown (Part 2/2): M\&A to Valuation. Metrics: \textbf{Accuracy (\%)} / \textbf{Score (0-5)}.}
\setlength{\tabcolsep}{12pt}
\resizebox{\linewidth}{!}{%
\begin{tabular}{l|cccccc}
\toprule
\textbf{Model} & \textbf{M\&A} & \textbf{Macro Mkt.} & \textbf{Macro Ind.} & \textbf{Port. Mgmt.} & \textbf{Risk Mgmt.} & \textbf{Valuation} \\ 
\midrule
 & \multicolumn{6}{c}{\textit{Proprietary Models}} \\ \midrule
Gemini-3.1 Pro & 66.3 / 4.1 & 50.2 / 4.0 & 49.1 / 4.2 & 58.2 / 4.0 & 68.4 / 4.2 & 75.5 / 4.2 \\
GPT-5.4 & 60.5 / 4.4 & 47.1 / 4.4 & 46.8 / 4.5 & 55.9 / 4.3 & 66.1 / 4.5 & 72.3 / 4.5 \\
Doubao-1.6 Vision & 57.4 / 4.4 & 45.3 / 4.0 & 41.5 / 4.3 & 52.4 / 4.0 & 62.3 / 4.1 & 68.7 / 4.2 \\
GPT-5 & 50.0 / 4.3 & 43.5 / 4.2 & 44.5 / 4.3 & 43.8 / 4.1 & 63.0 / 4.3 & 66.3 / 4.5 \\
Gemini-2.5 Pro & 52.1 / 4.0 & 39.5 / 3.9 & 39.4 / 4.0 & 41.5 / 3.6 & 55.9 / 4.0 & 58.2 / 4.0 \\
Gemini-2.5 Flash & 40.5 / 4.1 & 38.2 / 3.8 & 37.4 / 4.1 & 36.9 / 3.6 & 50.0 / 4.0 & 51.7 / 3.7 \\
Qwen3-Max & 15.8 / 3.8 & 24.0 / 3.5 & 27.0 / 3.8 & 23.6 / 3.1 & 35.1 / 3.7 & 24.5 / 3.5 \\
Grok-4.1 Fast & 22.6 / 3.5 & 25.7 / 3.2 & 27.6 / 3.6 & 22.5 / 2.8 & 35.3 / 3.3 & 24.0 / 3.0 \\ 
\midrule
 & \multicolumn{6}{c}{\textit{Open-Source Models}} \\ \midrule
Qwen3-VL-235B & 63.2 / 4.2 & 45.5 / 3.9 & 40.2 / 4.2 & 53.9 / 3.9 & 64.2 / 4.1 & 71.2 / 4.3 \\
Qwen3-VL-30B & 53.2 / 3.6 & 38.7 / 3.5 & 35.9 / 3.8 & 42.4 / 3.4 & 54.4 / 3.8 & 61.4 / 4.0 \\
Qwen2.5-VL-72B & 26.3 / 3.2 & 30.2 / 3.1 & 32.3 / 3.3 & 29.2 / 2.8 & 45.7 / 3.4 & 27.9 / 3.4 \\ 
\bottomrule
\end{tabular}%
}

\label{tab:domain_breakdown_part2}
\end{table*}
\clearpage      
\onecolumn      
\raggedbottom   
\setlength{\parindent}{0pt}  
\setlength{\parskip}{0.6em}  


\noindent\rule{\textwidth}{1.5pt}
\begin{center}
    \textbf{\large Prompt 1: Standard Model Inference Prompt}
\end{center}
\noindent\rule{\textwidth}{0.5pt}


\textbf{You are a professional financial domain expert.} I have uploaded a complete financial report PDF as page screenshots, along with a series of financial questions. These questions are guaranteed to be related to one or more charts within the PDF. Please answer all questions based on the relevant charts.

\textbf{The question types are:} single\_choice, multiple\_choice, numerical\_precise, numerical\_approximate, and open\_ended.

$\bullet$ For single\_choice, the answer is a single option letter. \\
$\bullet$ For multiple\_choice, the answer is multiple option letters. \\
$\bullet$ For numerical\_precise and numerical\_approximate, the answer is a pure number without any units. \\
$\bullet$ For open\_ended, the answer is a string of text.

\textbf{Your task is to determine the correct answer for each question.} \\
I will provide all questions at once. Please answer them in order. The response \textbf{MUST} be in JSON format.

\textbf{Output Requirements:}

1. Output \textbf{ONLY} the required JSON content, with no other explanatory text or information.

2. The format for each question's answer is a dictionary with the key "answer" and the value as a string.
\begin{itemize}[leftmargin=2em, nosep, label=$\circ$]
    \item Example for single\_choice: \{"answer": "C"\}
    \item Example for multiple\_choice: \{"answer": "ABD"\}
    \item Example for numerical\_precise/approximate: \{"answer": "12.3"\} (No units. Round or format decimals as the question requires).
    \item Example for open\_ended: \{"answer": "This is a text answer."\}
\end{itemize}

3. The outermost structure must be a list of these dictionary results, in the same order as the questions.

A strict reference for the output format is as follows (do not reference the content, only the format):

\begin{minipage}{\linewidth}
\begin{lstlisting}[
    basicstyle=\ttfamily\normalsize, 
    columns=fullflexible, 
    breaklines=true, 
    frame=single, 
    rulecolor=\color{black}, 
    language=json,
    aboveskip=0.5em,
    belowskip=0.5em
]
[
  { "answer": "C" },
  { "answer": "ABD" },
  { "answer": "12.3" },
  { "answer": "This is a text answer" }
]
\end{lstlisting}
\end{minipage}

Here is the set of financial questions to be answered: \textit{[Insert Questions List]}\\
\noindent\rule{\textwidth}{1.5pt}
\captionof{table}{Full evaluation prompts for Model Inference.}
\label{tab:eval_prompts1}

\clearpage 

\noindent\rule{\textwidth}{1.5pt}
\begin{center}
    \textbf{\large Prompt 2: LLM-as-a-Judge Scoring Prompt}
\end{center}
\noindent\rule{\textwidth}{0.5pt}


\textbf{You are a professional financial domain expert.} I will provide an open-ended financial question, its reference answer, and an answer to be evaluated. Please provide a score for the evaluated answer based on the question and the reference answer.

\textbf{Scoring Criteria:} \\
Scoring must be an integer between 0 and 5, where 0 is the lowest and 5 is the highest. The reference answer is for guidance only and does not necessarily represent a 5-point (perfect) answer.

\textbf{Output Format Requirement:} \\
Provide only a single integer between 0 and 5. Do not include any punctuation or other content.

\noindent\rule{0.3\textwidth}{0.5pt} \\ 
\textbf{[Input Template]}

The Question is: \textit{\{question\}}

The reference answer is: \textit{\{context\}}

The answer to be evaluated is: \textit{\{model\_answer\}} \\
\noindent\rule{\textwidth}{1.5pt} 
\captionof{table}{Full evaluation prompts for LLM-based Scoring.}
\label{tab:eval_prompts2}


\clearpage


\end{document}